 \documentclass[final,5p,times,twocolumn]{elsarticle}

\usepackage{amssymb}
\usepackage{amsmath}
\usepackage{bm}
\usepackage{afterpage}
\usepackage{enumerate}
\usepackage{algorithm}
\usepackage{algorithmicx}
\usepackage{algpseudocode}

\usepackage[pdftex,
colorlinks=true,
urlcolor=blue,
citecolor=blue,
linkcolor=blue,
hyperfootnotes=false]{hyperref}

\usepackage{tipa,tipx}

\newcommand{\HPhi}{ \mathcal{H} \Phi}

\def\ket#1{ | #1 \rangle }
\def\bra#1{ \langle #1 | }
\def\combin#1#2{{}^{~}_{#1}C^{~}_{#2}}
\def\nod{N^{~}_{\downarrow}}
\def\nodd{N^{\prime}_{\downarrow}}
\def\noi{N^{~}_{\rm xxz}}
\def\braket#1#2{ \langle #1 | #2 \rangle }
\def\repa{\mathfrak{a}}
\def\repb{\mathfrak{b}}
\def\syma{\mathsf{a}}
\def\symb{\mathsf{b}}

\newcounter{bla}

\journal{Computer Physics Communications}

\begin{document}

\begin{frontmatter}



\title{Quantum spin solver near saturation: QS$^3_{~}$}


\author[a,b,c]{Hiroshi Ueda\corref{author}}
\author[c,d,e,f]{Seiji Yunoki}
\author[g]{Tokuro Shimokawa}

\cortext[author] {Corresponding author.\\\textit{E-mail address:} h\_ueda@qiqb.osaka-u.ac.jp}
\address[a]{Center for Quantum Information and Quantum Biology, Osaka University,  Toyonaka, 560-0043, Japan.}
\address[b]{JST, PRESTO, Kawaguchi, 332-0012, Japan}
\address[c]{Computational Materials Science Research Team, 
RIKEN Center for Computational Science (R-CCS), Kobe, 650-0047, Japan}
\address[d]{Computational Condensed Matter Physics Laboratory, RIKEN Cluster for Pioneering Research (CPR), Wako, 351-0198, Japan}
\address[e]{Computational Quantum Matter Research Team, 
RIKEN Center for Emergent Matter Science (CEMS), Wako, 351-0198, Japan}
\address[f]{Quantum Computational Science Research Team, 
RIKEN Center for Quantum Computing (RQC), Wako, 351-0198, Japan}
\address[g]{Theory of Quantum Matter Unit, Okinawa Institute of Science and Technology Graduate University, Onna, 904-0495, Japan}

\begin{abstract}
We develop a program package named QS$^{3}$ [\textipa{kj\'u:-\'es-kj\'u:b}] based on the (thick-restart) Lanczos method for analyzing 
spin-1/2 XXZ-type quantum spin models on spatially uniform/non-uniform lattices near fully polarized states, which can be mapped to  
dilute hardcore Bose systems. 
All calculations in QS$^{3}$, including eigenvalue problems, expectation values for one/two-point spin operators, and 
static/dynamical spin structure factors,  are performed in the symmetry-adapted bases specified 
by the number $N_{\downarrow}$ of down spins and the wave number ${\bm k}$ associated with the translational symmetry 
without using the bit representation for specifying spin configurations. 
Because of these treatments, QS$^{3}$ can support large-scale quantum systems containing more than 1000 sites with 
dilute $N_{\downarrow}$. 
We show the benchmark results of QS$^{3}$ for the low-energy excitation dispersion of the isotropic Heisenberg model on the 
$10\times10\times10$ cubic lattice, 
the static and dynamical spin structure factors of the isotropic Heisenberg model on the $10\times10$ square lattice, 
and the open-MP parallelization efficiency on the supercomputer 
(Ohtaka) based on AMD Epyc 7702 installed at the Institute for the Solid State Physics (ISSP). 
Theoretical backgrounds and the user interface of QS$^{3}$ are also described. 

%
%
%
%

\end{abstract}



\end{frontmatter}



{\bf PROGRAM SUMMARY/NEW VERSION PROGRAM SUMMARY}

\begin{small}
\noindent
{\em Program Title:} QS$^{3}$ \\
{\em CPC Library link to program files:} (to be added by Technical Editor) \\
{\em Developer's respository link:} (if available) \\
{\em Code Ocean capsule:} (to be added by Technical Editor)\\
{\em Licensing provisions:} MIT \\
{\em Programming language:} Fortran90 \\
 {\em External routines/libraries:} 
 BLAS, LAPACK \\
{\em Nature of problem:}
Physical properties (such as total energy, magnetic moment, two-point spin correlation function, and dynamical structure factor) \\ 
{\em Solution method:} 
Application software based on the full diagonalization method and the exact diagonalization method using the Lanczos 
and thick-restart Lanczos techniques for quantum spin $S=1/2$ models such as the XXZ model. \\ 
{\em Restrictions:} Spin $S=1/2$ systems with U(1) symmetry. \\
{\em Unusual features:} Massively large quantum spin systems with U(1) symmetry near the saturation field can be solved with or without considering transnational symmetry, which is difficult to treat using standard exact diagonalization 
libraries with the bit representation for specifying spin configurations. 
   \\

\end{small}

\section{Introduction}
\label{Intro}

The exact diagonalization (ED) method~\cite{Dagotto94} is a traditional approach and indeed one of the most powerful numerical methods to correctly understand the nature of quantum many-body systems from finite-size cluster calculations.
This method enables us to obtain directly our target eigenvectors and the corresponding eigenvalues of Hamiltonian $\hat{H}$ and to evaluate all of the static, dynamical, and thermal properties without any bias by storing in physical memory of computers several vectors of the size as large as the dimension of the Hamiltonian matrix. 
Moreover, in contrast to a quantum Monte Carlo method, which in general suffers from the so-called negative sign problem for fermion and quantum frustrated spin systems, the ED method does not have such kind of difficulty. 
Therefore, the ED method has been used at the forefront in the research field of quantum many-body systems.

The only disadvantage of the ED method is that the accessible system size is severely limited to small clusters. 
A direct way to alleviate  
this disadvantage is to introduce the message-passing-interface (MPI) parallelization~\cite{MPI_Forum} and adapt the Hamiltonian symmetry~\cite{Sandvik2010}.
For examples, sophisticated ED program packages such as SpinPack~\cite{spinpack}, Rokko~\cite{rokko}, and $\HPhi$~\cite{Kawamura2017} are designed with MPI techniques for high performance on large-scale supercomputers and use U(1) and/or lattice symmetry to blockdiagonalize the Hamiltonian matrix.
Indeed, these large-scale computations with ingenious ways have led to new findings and many discoveries in the long history of studies in the quantum many-body systems, some of which can be found, e.g., in 
Refs.~\cite{Oitmaa1978,Leung1993, Waldtmann1998, Richter2010, Nakano2010, Nakano2011, Lauchli2011, Nakano2015, Nakano2019, Lauchli2019}.

Nonetheless, the accessible system sizes are still small because the required computational resources increase exponentially with the system size. 
For example, for a spin $S$=1/2 Heisenberg model without magnetic field, the conventional Lanczos algorithm~\cite{Lanczos:1950} can treat up to 50 spins by fully using the power of a large modern supercomputer~\cite{Wietek2018}. 
In some particular cases, the system size limitation to small clusters may not be an issue, e.g., in strong random systems~\cite{Watanabe2014, Kawamura2014, Shimokawa2015, Uematsu2017, Uematsu2018, Uematsu2019} 
and in the high-temperature limit, where the typical correlation lengths are usually small. 
However, the accessible system size becomes crucial when, for example, gapless or incommensurate phases are discussed in various intriguing quantum many-body systems.

Let us now remind a distinct advantage of the U(1) symmetry adaption for a quantum many-body system that can be mapped to a dilute particle system with a conserved number of particles, e.g., an $S=1/2$ XXZ model near the saturation field. 
In such a system, the ED method can treat much larger system sizes with $10^3$ sites and more, in principle, because the required matrix dimension is scaled as $O(N^{\nod}_{~})$, instead of $O(\exp(N))$, where $N$ is the system size and $\nod$ is the number of particles. 
A protocol in previous ED program packages~\cite{ spinpack, Kawamura2017, titpack, quspin1,quspin2} is to use a bitwise operation and thus the power of the U(1) symmetry is not fully exploited. 
While the bitwise operation is essential to speedup higher-level arithmetic operations, it requires that all basis states for expressing the Hamiltonian matrix are represented as binary numbers. 
Therefore, we cannot express the basis for more than 63 sites~\cite{note1} in a single binary number, e.g., when an $S=1/2$ spin model is treated on 64-bit processors.

In this paper, we develop an ED program package named QS$^3$ (Quantum Spin Solver near Saturation). 
This package can treat $S$=1/2 XXZ spin models near the saturation field where the U(1) symmetry is preserved with small $\nod =O(1)$. 
This is the first symmetry adapted open source ED package developed without using the bit representation. 
QS$^3$ also adapts the lattice symmetry of the Hamiltonian and thus it is able to treat $O(10^3)$ sites near the saturation field. 
A prototype of QS$^3$ has been already used for analyzing the ground state phase diagram of $S$=1/2 XXZ spin models 
on the triangular lattice near the saturation field with system sizes up to 1296 sites~\cite{Yamamoto2017}.

The QS$^3$ code is based on the Lanczos and thick-restart Lanczos methods~\cite{Wu1999,Wu2000} to calculate the low-energy 
eigenvalues and the corresponding eigenvectors of the Hamiltonian matrix, and is supported by the external libraries 
BLAS and LAPACK~\cite{anderson1999}. 
Available physical quantities are the local magnetization, two-point correlation functions, static spin structure factors, 
and dynamical spin structure factors. 
For the calculation of the dynamical spin structure factors, the continuous fractional expansion is employed with the Lanczos 
method~\cite{Dagotto94,Haydock72,Gagliano87}. 

The QS$^3$ code is specialized to the analysis for quantum magnets under a high magnetic field, dilute hard-core Bose gases, and low-energy properties of ferromagnets. Thus, it is particularly beneficial, for example, to the study of a field-induced spin nematic state~\cite{nematic0, nematic1, nematic2, nematic3,nematic4,nematic5,nematic6,nematic7,nematic8,nematic9,nematic10}, which often emerges in a high magnetic field. 
The QS$^3$ code can also be used to estimate magnetic couplings of effective spin Hamiltonians for some materials by directly comparing the spin excitations 
calculated numerically and measured experimentally by inelastic neutron scattering in a sufficiently high magnetic field. 
Therefore, the QS$^3$ package is useful for both theoretical and experimental researchers. 
The QS$^3$ package is designed to be executed on generally available computing resources such as laptops and small workstations, and therefore only OpenMP is used for parallelizations. 
The QS$^3$ package provides several samples for demonstration calculating physical quantities of $S$=1/2 XXZ models 
on three different lattices, the square, triangular, and cubic lattices, 
which preserve the translational symmetry. 
The QS$^3$ package also supports the analysis for systems without the translational symmetry. 

The rest of this paper is organized as follows. 
The basic usage of the QS$^3$ package is first described in Sec.~\ref{sec:basic_usage}.
The algorithms implemented in QS$^3$ are then explained in Sec.~\ref{sec:algorithms}.
In Sec.~\ref{sec:benchmark}, benchmark results on the square and cubic lattices are provided, and the bottlenecks and characteristics of the QS$^3$ calculations are discussed. 
Finally, the paper is summarized with brief discussion of future extension of the QS$^3$ package in Sec.~\ref{sec:summary}.

\section{Basic usage of QS$^{3}$}
\label{sec:basic_usage}

\subsection{How to download and build QS$^{3}$}
The QS$^3$ package, containing the Fortran source codes, samples, and manual, is available 
on GitHub (https://github.com/QS-Cube/ED). 
For building QS$^3$, Fortran compiler with BLAS/LAPACK library~\cite{anderson1999} is prerequisite. 

For those who have their own Git accounts, simply clone the repository on their local computers: 
\begin{quote}
\begin{itemize}
\$ git clone https://github.com/QS-Cube/ED.git
\end{itemize}
\end{quote}
Otherwise, go to the web page and click the ``Code" button and ``Download ZIP" to get ``ED-main.zip". 
The zip file is unpacked as 
\begin{quote}
\begin{itemize}
\$ unzip ED-main.zip \\
\$ cd ED-main
\end{itemize}
\end{quote}

A simple Makefile is provided to build the executable files ``QS3.exe" for systems preserving the translational symmetry and ``QS3\_only\_u1.exe" for systems without the translational symmetry. 
The following procedures after the cloning or downloading will generate the executable file and execute sample programs
\begin{quote}
\begin{itemize}
\$ cd script \\
\$ ./make.sh
\end{itemize}
\end{quote}
After executing samples, each result is stored in the separate directory, "output\_ex1", "output\_ex2", $\cdots$, and "output\_ex5". 

\subsection{Model}
The QS$^3$ package can treat the following $S$=1/2 XXZ-type spin Hamiltonian: 
\begin{equation}
\hat{\mathcal{H}} = \sum_{r < r'} \{ J_{r,r'}^{xy} ( \hat{s}_r^x \hat{s}_{r'}^x+ \hat{s}_r^y \hat{s}_{r'}^y) 
+ J_{r,r'}^z  \hat{s}_r^z \hat{s}_{r'}^z \} -  h^z \sum_{r=1}^{N} \hat{s}_r^z,
\label{eq:xxz_hamiltonian}
\end{equation}
where $\hat{{\bm s}}_r=(\hat{s}^x_r,\hat{s}^y_r,\hat{s}^z_r)$ is a spin $S=1/2$ operator at site $r$ on an $N$-site cluster, $J_{r,r'}^{z\, (xy)}$ is the $z\,(xy)$ component of the two-body exchange interaction between the $r$th and $r'$th spins, and $h^z$ is the uniform magnetic field applied along the $z$ direction.
Since the Hamiltonian $\hat{\mathcal{H}}$ commutes with the $z$ component of the total spin, i.e., 
$[\hat{\mathcal{H}},\sum_r \hat{s}^z_r]=0$, the U(1) symmetry is preserved. 
We also assume that the Hamiltonian $\hat{\mathcal{H}}$ is translationally invariant under periodic boundary conditions. 
In the QS$^3$ package, the lattice structures and the range of the exchange interactions can be varied 
as long as the U(1) and translational symmetries are preserved. 
The details will be described below. 

\subsection{How to use QS$^{3}$}
\begin{figure*}[t]
 \begin{center}
  \includegraphics[width=19.0cm]{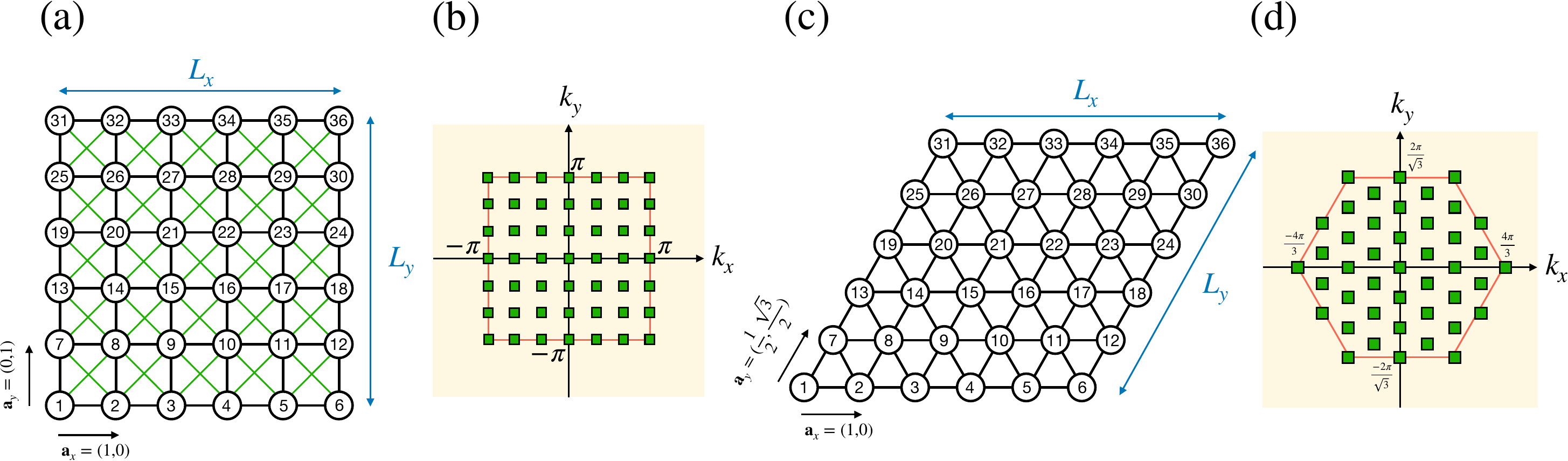}
  \caption{
  Examples of lattice structures and the corresponding momentum spaces. 
  (a, b) The 6 $\times$ 6 square lattice cluster and (c, d) the 6 $\times$ 6 triangular lattice cluster.  
  The numbers in each circle in (a) and (c) denote site numbers used in the input file. ${\bf a}_x$ and ${\bf a}_y$ represent the primitive translational vectors, where the lattice constant is set to be one. 
  $L_x$ ($L_y$) is the linear dimension of the lattice along the ${\bf a}_x$ (${\bf a}_y$) direction (in these examples, $L_x=L_y=6$). 
  In (b) and (d), the red lines indicate the first Brillouin Zone, and the green squares represent the allowed momentum points, ${\bf k}=(k_x, k_y)$, for the clusters in (a) and (c), respectively, under periodic boundary conditions. 
  The QS$^3$ package can compute eigenvalues and eigenvectors of the Hamiltonian matrix at each momentum sector, separately. 
  These momentum points correspond to wave vector points, ${\bf q}=(q_x, q_y)$, which can be chosen for computing the dynamical spin structure factor ${\bf S}({\bf q}, \omega)$.}
  \label{fig:lattice}
 \end{center}
\end{figure*}

Here we explain in detail how to use QS$^{3}$ by providing a concrete and simple example of the $S=1/2$ isotropic Heisenberg model, 
i.e., $J_{r,r'}^{xy}=J_{r,r'}^z=J_1$ when sites $r$ and $r'$ are nearest neighbored and $J_{r,r'}^{xy}=J_{r,r'}^z=0$ otherwise, 
on the 6$\times$6 square lattice 
shown in Fig.~\ref{fig:lattice}(a). 
We calculate the 10 lowest eigenvalues and the corresponding eigenvectors in the subspace of $\langle \sum_r \hat{s}^z_r \rangle=15$ and momentum ${\bf k}=(k_x, k_y)=(0, 0)$ by using the thick-restart Lanczos algorithm. 
Here, $\langle \sum_r \hat{s}^z_r \rangle$ is the expectation value of the $z$ component of the total spin with respect to an eigenvector. 
We also calculate the local magnetization and the two-point correlation function from the obtained eigenvectors, and the dynamical spin structure factor $S^+({\bf q}, \omega)$ at wave vector ${\bf q}=(q_x, q_y)=(0, 0)$ by means of the continued fraction method. 

\subsubsection{Set a main input file}
One should first create several input files in the ``input\_ex1'' directory.
A concrete example of the main input file, input.dat,  for the $S$=1/2 isotropic Heisenberg model on the 
6$\times$6 square lattice is shown below. 
\begin{quote}
 \begin{itemize}
${\rm \& input \_parameters}$ \\
${\rm \ \ NOS    = 36, }$ \\
${\rm \ \ NOD    = 3, }$ \\
${\rm \ \ LX     = 6, }$ \\
${\rm \ \ LY     = 6, }$ \\
${\rm \ \ LZ     = 1, }$ \\
${\rm \ \ KX     = 0, }$ \\
${\rm \ \ KY     = 0, }$ \\
${\rm \ \ KZ     = 0, }$ \\
${\rm \ \ NOxxz  = 72, }$ \\
${\rm \ \ ALG=2, }$ \\
${\rm \ \ cal\_lm = 1, }$ \\%
${\rm \ \ cal\_cf = 1, }$ \\%
${\rm \ \ cal\_dsf= 1, }$ \\%
${\rm \ \ wr \_wf  = 1, }$ \\
${\rm \ \ re \_wf  = 0, }$ \\
${\rm \ \ FILExxz= ``input\_ex1/list \_ xxz \_ term \_ 36.dat", }$ \\
${\rm \ \ FILEwf = ``work/", }$ \\
${\rm \ \ OUTDIR = ``output\_ex1/", }$ \\
${\rm  \& end }$

${\rm \& input \_static }$ \\
${\rm \ \ NOV   = 2, }$ \\
${\rm \ \ NOLM   = 36, }$ \\
${\rm \ \ NOCF   = 1296, }$ \\
${\rm \ \ FILElm = ``input\_ex1/list \_ local \_ mag.dat", }$ \\
${\rm \ \ FILECF = ``input\_ex1/list \_ cf \_ ss.dat", }$ \\
${\rm  \& end}$ 

${\rm \& input \_dynamic }$ \\
${\rm \ \ spsmsz = 1, }$ \\
${\rm \ \ itr \_dsf= 200, }$ \\
${\rm \ \ QX     = 0.0d0, }$ \\
${\rm \ \ QY     = 0.0d0, }$ \\
${\rm \ \ QZ     = 0.0d0, }$ \\
${\rm \ \ rfield = 0.495d0, }$ \\
${\rm \ \ FILEpos= ``input\_ex1/list \_ site \_ position \_36 \_type1.dat", }$ \\
${\rm  \& end}$ 

${\rm \& input \_lancz }$ \\ %
${\rm \ \ lnc \_ene \_conv0 = 1.0d{\rm \mathchar`-}14, }$ \\%
${\rm \ \ minitr=20, }$ \\%
${\rm \ \ maxitr = 10000, }$ \\%
${\rm \ \ itrint = 5, }$ \\%
${\rm  \& end }$ %

${\rm \& input \_TRLan }$ \\ 
${\rm \ \ NOE = 10, }$ \\
${\rm \ \ NOK = 15, }$ \\
${\rm \ \ NOM = 30, }$ \\
${\rm \ \ maxitr = 10000, }$ \\
${\rm \ \ lnc \_ene \_conv = 1.0d{\rm \mathchar`-}14, }$ \\
${\rm \ \ i\_vec\_min = 1, }$ \\
${\rm \ \ i\_vec\_max = 1, }$ \\
${\rm  \& end}$ 

 \end{itemize}
\end{quote}
The main input file given above consists of five parts, ${\rm input \_parameters}$, ${\rm  input \_static }$, ${\rm input \_dynamic }$, 
${\rm input \_lancz }$, and ${\rm input\_TRLan}$. 
The meaning of each part and the variables used there are explained below.\\

\begin{description}
\item{\underline{input\_parameters}} 

This part requires the users to set fundamental conditions, number of spins, number of down spins, linear dimensions of the cluster, momentum sector, location of the input file specifying the lattice structure and the exchange interactions, locations of outputs for the results, and an algorithm for the calculation of eigenvalues/eigenvectors. The details of the variables are explained below.

{\bf NOS} (INTEGER): Number $N$ of spins. \\
{\bf NOD} (INTEGER): Number $\nod$ of down spins. The users can select the $z$ component of the total spin, 
$M=(N/2-\nod)$, by adjusting this valuable. \\
{\bf LX, LY, LZ} (INTEGER): Linear dimensions $L_x$, $L_y$, and $L_z$ of the cluster in the $x$, $y$, and $z$ directions. \\
{\bf KX, KY, KZ} (INTEGER): Momentum sector $K_x$, $K_y$, and $K_z$. The users should set a allowed momentum value that is compatible with 
the cluster size and shape, ${\bf k} \cdot {\bf a}_\alpha = 2\pi K_\alpha/L_\alpha$ with $\alpha \in \{ x, y, z\}$, 
where ${\bf k}$ is the momentum and ${\bf a}_\alpha$ is the $\alpha$ component of the primitive translation vectors. \\
{\bf NOxxz} (INTEGER): Number $N^{~}_{\rm xxz}$ of the two-body exchange interactions. \\
{\bf ALG} (INTEGER): The users can choose algorithms by setting 1: Conventional Lanczos, 2: Thick-restart Lanczos, 3: Full diagonalization. \\
{\bf cal$\_$lm, cal$\_$cf, cal$\_$dsf} (INTEGER): The local magnetization (lm), spin correlation function (cf), and dynamical structure factor (dsf) are calculated by setting the corresponding variables to 1. Otherwise, these variables should be 0.\\
{\bf wr$\_$wf} (INTEGER): If ${\rm wr\_wf}=1$, the computed eigenvectors are output in the directory specified by {FILEwf}. \\
{\bf re$\_$wf} (INTEGER): If ${\rm re\_wf}=0$, the whole diagonalization calculation starts from a random initial vector. 
If ${\rm re\_wf}=1$, only the expectation values of physical quantities are computed 
after reading the eigenvectors already prepared in the directory 
specified by  {FILEwf}. \\
{\bf FILExxz} (CHARACTER): The location of the file that defines the lattice structure and the two-body exchange interactions 
$J_{r,r'}^z$ and $J_{r,r'}^{xy}$. The lattice structure is specified in terms of pairs of sites connected by the nonzero interactions. 
The details are described below.\\
{\bf FILEwf} (CHARACTER): The location of output for the computed eigenvectors. The number of eigenvectors is specified 
by NOE (the number of the lowest eigenvalues/eigenvectors calculated by an eigensolver specified by ALG).\\
{\bf OUTDIR} (CHARACTER): The location of output for the computed expectation values such as the local magnetization, spin correlation function, and dynamical structure factor. \\
\end{description}

\begin{description}
\item{\underline{input\_static } }

This part requires the users to set the conditions for the calculation of the static physical quantities, i.e., the local magnetization and two-point correlation function. The users should prepare separately the two files that specify the sites and the pairs of sites for the calculation of the local magnetization and the two-point correlation function, respectively.\\

{\bf NOV} (INTEGER): Number of the lowest eigenvectors used for computing static physical quantities.\\
{\bf NOLM} (INTEGER): Number of sites for which the local magnetization is computed. If NOLM=0, the local magnetization is not computed.\\
{\bf NOCF} (INTEGER):  Number of pairs of sites for which the two-point correlation function is computed. If NOCF=0, the correlation function is not computed.\\
{\bf FILElm} (CHARACTER): The location of the file specifying the site definition of the cluster for the calculation of the local magnetization. The details of the file are explained below. \\
{\bf FILECF} (CHARACTER): The location of the file specifying the pairs of sites for the calculation of the two-point correlation function. The details of the file are explained below.  \\
\end{description}

\begin{description}
\item{\underline{input\_dynamic} }

This part requires the users to set the conditions for the calculation of the dynamical spin structure factor. 
The user should prepare the file specifying the sites of the cluster, separately.\\

{\bf spsmsz} (INTEGER): $S^+({\bf q}, \omega)$,  $S^-({\bf q}, \omega)$, and $S^z({\bf q}, \omega)$ 
are computed by setting the value ${\rm spsmsz}=1$, $2$, and $3$, respectively. 
If ${\rm spsmsz}=0$, the dynamical spin structure factor is not computed. \\
{\bf itr$\_$dsf} (INTEGER): Number of iterations for the continued fraction method. See Sec.~\ref{sec:continued_fraction_expansion} for the details. \\
{\bf QX,QY,QZ} (REAL8): Wave vector point ${\bf q}$ at which the dynamical spin structure factor is computed, ${\bf q} \cdot {\bf a}_\alpha = Q_\alpha$. \\
{\bf rfield} (REAL8): Magnetic field value $h_z$. \\
{\bf FILEpos} (CHARACTER): The location of the file specifying the site positions. The details of the file are explained below.  \\
\end{description}

\begin{description}
\item{\underline{input\_lancz} }

This part requires the users to set the conditions for the conventional Lanczos algorithm when the users set ${\rm ALG}=1$. 
The users must set ${\rm NOE}=1$ below because only the lowest eigenvalue with the corresponding eigenvector is computed here. 
\\

{\bf lnc$\_$ene$\_$conv0} (REAL8): Convergence condition for the Lanczos iteration.  \\
{\bf min(max)itr} (INTEGER): The minimum/maximum number of iterations for the Lanczos method.  \\
{\bf itrint} (INTEGER): Every itrint iterations, the convergence of the Lanczos iteration is checked.  \\
\end{description}

\begin{description}
\item{\underline{input\_TRLan } }

This part requires the users to set the conditions for the thick-restart Lanczos algorithm when the users set ${\rm ALG}=2$.
\\

{\bf NOE} (INTEGER): Number of the lowest eigenvalues/eigenvectors computed by the thick-restart Lanczos method. \\
{\bf NOK} (INTEGER):  $N_{\rm K}$ value. See Algorithm \ref{algorithm:TRL}. \\
{\bf NOM} (INTEGER):  $N_{\rm M}$ value. See Algorithm \ref{algorithm:TRL}. \\
{\bf maxitr} (INTEGER): $I_{\rm M}$ value. See Algorithm \ref{algorithm:TRL}. The maximum number of iterations for the thick-restart Lanczos method.  \\
{\bf lnc$\_$ene$\_$conv} (REAL8): Convergence condition for the thick-restart Lanczos iteration.  \\
{\bf i$\_$vec$\_$min/max} (INTEGER): Store the {i$\_$vec$\_$min}-th to {i$\_$vec$\_$max}-th lowest 
eigenvectors computed by the thick-restart Lanczos method in the location specified by FILEwf. \\
\end{description}

\subsubsection{Set a file specifying the model}
The users are required to specify the lattice structure and the exchange interactions in an input file. 
We provide  in the input directory an example, 
list$\_$xxz$\_$term$\_$36.dat, for the $S$=1/2 isotropic Heisenberg model on the 6$\times$6 square lattice. 
The file location should be specified in the main input file using the FILExxz variable. 
The first part of list$\_$xxz$\_$term$\_$36.dat are shown below.\\
\begin{quote}
 \begin{itemize}
   1 \ \ \ 2 \ $-1.0$E+00 \ $-1.0$E+00 \\
   2 \ \ \ 3 \ $-1.0$E+00 \ $-1.0$E+00 \\ 
   3 \ \ \ 4 \ $-1.0$E+00 \ $-1.0$E+00 \\
   4 \ \ \ 5 \ $-1.0$E+00 \ $-1.0$E+00 \\

.... \\
 \end{itemize}
\end{quote}
In this data statement, the first and second columns denote the pair of sites ($r$ and $r'$), and the third and fourth columns represent the corresponding exchange interactions, $J_{r,r'}^{xy}$ and $J_{r,r'}^{z}$, respectively [see Fig.~\ref{fig:lattice}].

\subsubsection{Set files for computing physical quantities}
The users are required to set several input files for the calculation of physical quantities. 
For the local magnetization $\langle {\hat s}_r^z \rangle$, 
the users have to specify the site $r$ for which the local magnetization is computed. 
For the two-point correlation function $\langle {\hat s}_r^{+} {\hat s}_{r'}^{-} \rangle$ 
and $\langle {\hat s}_{r}^z {\hat s}_{r'}^z \rangle$, 
the users have to specify the pair of sites ($r$, $r'$) for which the correlation function is computed. 
As concrete examples, we provide two files in the input directory, 
list$\_$local$\_$mag.dat and list$\_$cf$\_$ss.dat, for the local magnetization and the two-point correlation function, respectively. 

For the dynamical spin structure factor, the users are required to specify the site positions in the cluster. 
We also provide an example input file, list$\_$site$\_$position$\_$36$\_$type1.dat.
Note that the lattice constant is set to be one [see Fig.~\ref{fig:lattice}(a)].
The location of these files is specified using FILExxz, FILElm, and FILECF variables in the main input file.

\subsubsection{Run and results}
After preparing all these files described above, the users can perform the calculation as follows: \\

\$ {./QS3.exe $<$ input\_ex1/input.dat $>>$ output\_ex1/output.dat 2$>$$\&$1} \\

\noindent
Here, output.dat is the result file for the calculation, from which the users can check the status of the calculation.
The computed results of the physical quantities are output in the directory specified by the OUTDIR variable in the main input file. 
The eigenvectors are output in the directory specified by the FILEwf variable in the main input file.

\section{Implemented algorithms}
\label{sec:algorithms}

\subsection{Representation of states with the $U(1)$ symmetry}
The QS$^3$ package diagonalizes the Hamiltonian matrix for $\hat{\mathcal{H}}$ given in Eq.~(\ref{eq:xxz_hamiltonian}), 
consisting of $N$ spins that can be as large as $\mathcal{O}(10^3)$, with a small number $\nod = \mathcal{O}(1) \ll N$ of down spins 
or equivalently with a large total magnetization value $M \equiv \bra{\phi} \sum_{r}^{~} \hat{s}^z_r \ket{\phi} = N/2-\nod$, 
where $\ket{\phi}$ is an eigenstate of $\hat{\mathcal{H}}$. 

A standard way to construct the spin basis states, and accordingly represent the Hamiltonian matrix, is to use the bit representation, where up ($\uparrow$) and down ($\downarrow$) spins are expressed as 0- and 1-bit values, respectively. 
However, this is not a practical way for our purpose because a single four-byte (eight-byte) integer can only represent spin basis states up to $N$=31 (63)~\cite{note1} in a standard 32 (64)-bit operating system. 
Furthermore, the definition and operation of arbitrary-byte integers are not supported in standard numerical programming languages. 

Let us now explain how to construct the spin basis states in the QS$^3$ package. 
We first introduce the following fully polarized state as a vacuum state:
\begin{equation}
\ket{{\rm v}} \equiv \ket{\overbrace{\uparrow \uparrow \cdots \uparrow}^{N~{\rm spins}}}. 
\end{equation}
Each spin basis state $|a\rangle$ is then constructed by acting the $S=1/2$ descending operator $\hat{s}^{-}_r$ on the vacuum state, 
i.e., 
\begin{equation}
| a \rangle=\prod_{m=1}^{\nod} \hat{s}^-_{r_m} \ket{{\rm v}},
\label{eq:reference}
\end{equation}
where ${r_m}$ is the position of the $m$th down spin in real space. In the QS$^3$ package, 
a set of $\{ r_m \}^{~}_{1\leq m \leq \nod}$ is stored in $\nod$-dimensional integer vector (array) 
${\bf n}
\equiv(n_1, n_2, ..., n_{\nod})
=(r_1, r_2, ..., r_{\nod})
$ 
in ascending order, 
$1\leq r_1 < r_2 < \cdots < r_{\nod} \leq N$, implying that $r_m\geq m$.

We now focus on a subspace of the entire Hilbert space of $\hat{\mathcal{H}}$ by setting the number $\nod$ of down spins. 
The dimension of the subspace is $\combin{N}{\nod}$ and the spin basis states $|a\rangle$ in this subspace are numerated as 
$a=1,2,\cdots, \combin{N}{\nod}$. 
For a given set of $\{ r_m \}^{~}_{1\leq m \leq \nod}$, we can define an integer index $a$ through the following bijective function $F$: 
\begin{equation}
a=F({\bm n});~F({\bm n})= 1 + \sum_{m=1}^{\nod} {}^{~}_{r^{~}_m-1}C^{~}_{m}.
\label{eq:a_of_n}
\end{equation}
This one-to-one correspondence between $a$ and $\{ r_m \}^{~}_{1\leq m \leq \nod}$ can be understood as follows: 
When the $\nod$th down spin is located at 
$r^{~}_{\nod}$th site, the target spin configuration $\ket{a}$ should be listed after 
${}^{~}_{r^{~}_{\nod}-1}C^{~}_{\nod}$ patterns for arranging $\nod$ spins 
stored within sites from the first site to the ($r^{~}_{\nod}-1$)-th site.
We can apply the same procedure for $m=\nod-1$ down to $m=1$, recursively, 
assuming that $\combin{k}{m}=0$ if $k<m$. 

The inverse bijective function ${\bm n}=\bar{F}(a)$ is given in Algorithm~\ref{algorithm:f_bar} with the binary search algorithm 
(Algorithm~\ref{algorithm:binary_search}). 
It is highly instructive to first consider a concrete example. 
For example, let us consider the case of $N=8$, $\nod=4$, and ${\bm n}=(2,4,6,8)$, which corresponds to 
$a={}^{~}_{7}C^{~}_{4} + {}^{~}_{5}C^{~}_{3} + {}^{~}_{3}C^{~}_{2} + {}^{~}_{1}C ^{~}_{1} + 1 = 50$, 
according to Eq.~(\ref{eq:a_of_n}). 
Now, giving $s=a=50$ as the input, we explain how Algorithm~\ref{algorithm:f_bar} outputs ${\bm n}=(2,4,6,8)$.
The algorithm first searches for $n^{~}_{4}=r^{~}_{4}=8$ that satisfies 
${}^{~}_{7}C^{~}_{4}=35 < s = 50 \leq {}^{~}_{8}C^{~}_{4}=70$ and updates $s:=s-35=15$. 
Second, the algorithm searches for $n^{~}_{3}=r^{~}_{3}=6$ that satisfies
${}^{~}_{5}C^{~}_{3}=10 < s = 15 \leq {}^{~}_{6}C^{~}_{3}=20$ and updates $s:=s-10=5$. 
Third, the algorithm searches for $n^{~}_{2}=r^{~}_{2}=4$ that satisfies
${}^{~}_{3}C^{~}_{2}=3 < s = 5 \leq {}^{~}_{4}C^{~}_{2}=6$ and updates $s:=s-3=2$. 
Finally, the algorithm assigns $n^{~}_{1}=r^{~}_{1}=s=2$. 

More generally, Algorithm~\ref{algorithm:f_bar} first searches for $n^{~}_{\nod}=r^{~}_{\nod}$ that satisfies 
${}^{~}_{r^{~}_{r^{~}_{\nod}}-1}C^{~}_{\nod} < a \leq {}^{~}_{r^{~}_{\nod}}C^{~}_{\nod}$, 
followed by a research for $n^{~}_{\nod-1}=r^{~}_{\nod-1}$ that satisfies 
${}^{~}_{r^{~}_{\nod-1}-1}C^{~}_{\nod-1} < a-{}^{~}_{r^{~}_{\nod}-1}C^{~}_{\nod} \leq {}^{~}_{r^{~}_{\nod-1}}C^{~}_{\nod-1}$, 
until a search for $n^{~}_{2}=r^{~}_{2}$ that satisfies 
${}^{~}_{r^{~}_{2}-1}C^{~}_{2}< a-\sum^{\nod}_{m=3}{}^{~}_{r^{~}_{m}-1}C^{~}_{m} \leq {}^{~}_{r^{~}_{2}}C^{~}_{2}$. 
Finally, it uses Eq.~(\ref{eq:a_of_n}) to determine $n^{~}_{1}=r^{~}_{1}=a-\sum^{\nod}_{m=2}{}^{~}_{r^{~}_{m}-1}C^{~}_{m}$ 
and returns $\bm{n}$.
Note that the memory cost with $O(\nod\combin{N}{\nod})$ bytes for keeping the basis sets $\{{\bm n}\}$ in a computer, which may 
become a memory bottleneck for the Lanczos method, can be reduced 
to $O(1)$ by representing the spin basis states with $|a\rangle$ 
at the expense of additional 
numerical cost of $O(\nod \ln (N-\nod))$ 
for the use of the function $\bar{F}$.

\begin{figure}[h]
\begin{algorithm}[H]
  \caption{Generate ${\bm n}$ for given $a$.} 
  \label{algorithm:f_bar}
   \begin{algorithmic}[1]
    \Require{integers $\nod$, $N$, and $a$ with $1 \leq a \leq \combin{N}{\nod}$.}
    \Ensure{$\nod$-dimensional integer vector ${\bm n}$ with $1\leq n^{~}_1 < n^{~}_2 < \cdots < n^{~}_{\nod} \leq N$.}
    \Function{f\_bar}{$a$, $\nod$, $N$}
    \State $s:=a$
    \State $j:=N$
    \For{$m = \nod$ to 2 with $m:=m-1$}
    	\State $(j_0,f)=${\sc binary\_search}$(s-1,\{\combin{k}{m} \}_{1 \leq k \leq j},m,j)$
		\Statex \Comment A logical parameter $f$ is not used in this function.
		\Statex \Comment Note that $\combin{k}{m}=0$ if $k<m$.
		\State $j := j_0$
		\State $n^{~}_{m}:=j_0+1$
		\State $s:=s-\combin{j}{m}$
    \EndFor
    \State $n^{~}_1:=s$
    \State \Return(${\bm n}$)
    \EndFunction
   \end{algorithmic}
\end{algorithm}
\end{figure}
\begin{figure}[h]
\begin{algorithm}[H]
  \caption{Check whether $s_0$ is in ${\bm s}$.}
  \label{algorithm:binary_search}
   \begin{algorithmic}[1]
    \Require{integer $s_0$, $l_{\rm s}$, and $l_{\rm e}\,(\geq l_{\rm s})$, and $l_{\rm e}$-dimensional integer vector ${\bm s}$.}
    \Ensure{integer $p$ and logical $f$.}
    \Function{binary\_search}{$s_0,{\bm s},l_{\rm s},l_{\rm e}$}
    \State Search integer $p$ in $\{s_k\}^{~}_{l_{\rm s} \leq k \leq l_{\rm e}}$ satisfying $s_p \leq s_0 < s_{p+1}$ with $l_{\rm s} \leq p \leq l_{\rm e}$ by the binary search where $s^{~}_{l_{\rm e}+1}=\infty$. 
    \State $f := \left\{ \begin{matrix} 
    {\rm True} & s_p = s_0 \\ 
    {\rm False} & {\rm otherwise}
    \end{matrix} \right.$ 
    \State \Return ($p$, $f$)
    \EndFunction
   \end{algorithmic}
\end{algorithm}
\end{figure}
%


\subsection{Generation of the Hamiltonian matrix}
The Hamiltonian matrix is block diagonal with respect to the number $\nod$ of down spins and the dimension of the 
block-diagonal matrix specified with $(N,\nod)$ is $\combin{N}{\nod}$. 
There is a nonzero diagonal contribution of $\bra{a} \hat{s}^z_r \hat{s}^z_{r'} \ket{a} = \pm 1/4$ to the Hamiltonian matrix,  
where the sign of the value is minus when either $r$ or $r'_{~}$ is in ${\bm n}$ and otherwise it is plus. 
For this check, the QS$^3$ package uses Algorithm~\ref{algorithm:binary_search} of the binary search.
On the other hand, the off-diagonal operator $(\hat{s}^+_r \hat{s}^-_{r'} + \hat{s}^-_r \hat{s}^+_{r'})$ acting on a state $\ket{a}$ 
can generate a different basis state $\ket{a'}$ and the corresponding array ${\bm n}'=\bar{F}(a')$. 
We can also use the binary search to determine whether a new state is generated. 
Namely, the new state $|a'\rangle$ is generated when two logical variables $f^{~}_r$ and $f^{~}_{r'}$, which are given 
by $(p^{~}_t,f^{~}_t):=${\sc binary\_search}$(t,{\bm n},1,\nod)$ with $t \in \{r,r'\}$, are different. 
If this is the case, the QS$^3$ package uses Algorithm~\ref{algorithm:spin_exchange} to make the new array ${\bm n}'$ 
and we can obtain the off-diagonal matrix element $\bra{a'} \hat{s}^+_r \hat{s}^-_{r'} + \hat{s}^-_r \hat{s}^+_{r'} \ket{a} = 1$.
To construct the full matrix elements, we have to consider all sets of $\{r, r'\}$ compatible with the nonzero exchange interactions 
in $\hat{\mathcal{H}}$ given in Eq.~(\ref{eq:xxz_hamiltonian}). 
For this purpose, the QS$^3$ package uses Algorithm~\ref{algorithm:full_hamiltonian}, where the $N_{\rm xxz}$ variable is 
the number of the exchange 
interactions, i.e., the number of pairs $\{r, r'\}$ connected via the nonzero exchange interactions, and should be equal to 
NOxxz in the input file. 
Note that the contribution of the Zeeman term in Eq.~(\ref{eq:xxz_hamiltonian}) is excluded in 
Algorithm~\ref{algorithm:full_hamiltonian} because it is simply constant within the subspace of a fixed $\nod$.

\begin{figure}[h]
\begin{algorithm}[H]
  \caption{Spin exchange interaction between sites $r$ and $r'$.}
  \label{algorithm:spin_exchange}
   \begin{algorithmic}[1]
    \Require{integer $r$, $p$, $r'\,(>r)$, and $p'\,(\geq p)$, $\nod$-dimensional integer vector ${\bm n}$, and logical $f$, where $p$, $p'$, and $f$ are given by $(p,f):=${\sc binary\_search}$(r,{\bm n},1,\nod)$ and $(p',f'):=${\sc binary\_search}$(r',{\bm n},1,\nod)$, 
    assuming that $f'\ne f$.}
    \Ensure{$\nod$-dimensional integer vector ${\bm n}'$ with $1 \leq n'_1 < n'_2 < \cdots < n^{\prime}_{\nod} \leq N$.}
    \Function{spin\_exchange}{$r,p,f,r',p',{\bm n}$}
    	\If {$f = {\rm False}$}
		\State ${\bm n}':=(n_1,\cdots,n_{p},r,n_{p+1},\cdots,n_{p'-1},n_{p'+1} \cdots,n_{\nod})$
	\Else
		\State ${\bm n}':=(n_1,\cdots,n_{p-1},n_{p+1},\cdots,n_{p'},r,n_{p'+1} \cdots,n_{\nod})$	
	\EndIf
	\State \Return(${\bm n}'$)
    \EndFunction
   \end{algorithmic}
\end{algorithm}
\end{figure}
\begin{figure}[h]
\begin{algorithm}[H]
  \caption{Generation of full Hamiltonian matrix.}
  \label{algorithm:full_hamiltonian}
   \begin{algorithmic}[1]
    \Require{$\noi$-dimensional integer vectors ${\bm r}$ and ${\bm r}'$, and $\noi$-dimensional real vectors ${\bm J}^{xy}_{~}$ and ${\bm J}^{z}_{~}$.}
    \Ensure{$\combin{N}{\nod}$-dimensional real matrix ${\bf H}=\{ h_{a,a'}\}$.}
    \Function{gen\_full\_ham}{${\bm r},{\bm r}',{\bm J}^{xy}_{~},{\bm J}^{z}_{~}$}
    	\State ${\bf H}:=0$    
	\For {$a = 1$ to $\combin{N}{\nod}$}
		\State ${\bm n}:=${\sc f\_bar}$(a,\nod,N)$
		\For {$n = 1$ to $\noi$}
			\State $(p,f):=${\sc binary\_search}$(r^{~}_n,{\bm n},1,\nod)$
			\State $(p',f'):=${\sc binary\_search}$(r'_n,{\bm n},1,\nod)$
			\If {$f = f'$}
				\State $h^{~}_{a,a} := h^{~}_{a,a}  + J^z_n/4$
			\Else
				\State $h^{~}_{a,a} := h^{~}_{a,a} - J^z_n/4$			
				\State ${\bm n}':=${\sc spin\_exchange}$(r,p,f,r',p',{\bm n})$
				\State $a':=F({\bm n}')$
				\State $h^{~}_{a,a'} := J^{xy}_n/2$
			\EndIf
		\EndFor
	\EndFor
	\State \Return ({\bf H})
    \EndFunction
   \end{algorithmic}
\end{algorithm}
\end{figure}


\subsection{Representative states and Hamiltonian matrix elements in symmetry-adapted basis sets}
Not only the U(1) symmetry in spin space, but also lattice symmetry such as translational symmetry and point group symmetry 
can be used to reduce the dimension of the Hamiltonian matrix to be diagonalized and thus the computational cost. 
Here, we describe how to block-diagonalize the Hamiltonian $\hat{\mathcal{H}}$ based on the symmetry-adapted basis sets.

First, we briefly explain how to construct the symmetry-adapted basis sets that are the eigenstates of the lattice translational operator $\hat{T}$. 
A pedagogical introduction for the construction of the symmetry-adapted basis sets can be found in Ref.~\cite{Sandvik2010}. 
For simplicity, we consider a periodic chain with $N$ spins, namely $(L_x,L_y,L_z)=(N,1,1)$, in which the translational operator $\hat{T}$ is defined by 
shifting the position of the spin one site right, $\hat{T}\ket{a}=\prod_{m=1}^{\nod} \hat{s}^-_{r_m+1} \ket{{\rm v}}$, with $\hat{s}_{L+1}=\hat{s}_{1}$ under periodic boundary conditions. 
Note that the translational operator $\hat{T}$ is commutable with the Hamiltonian, i.e., $[\hat{T},\hat{H}]=0$, and 
the accessible eigenvalues of $\hat{T}$ are given as $\{ e^{ik}_{~}~|~k=2\pi K/N,\, 0 \leq K < N \}$ 
with momentum $k$ or momentum sector $K$.

The symmetry-adapted basis states with a given momentum $k$ is given as 
\begin{equation}
\ket{a,k}=\frac{1}{\sqrt{N^{~}_{a,k}}}
\sum_{j=1}^{N}
e^{-ikj}_{~} \hat{T}^{j} \ket{a}, 
\label{eq:akstate}
\end{equation}
where $|a\rangle$ is a single reference state with a fixed number $\nod$ of down spins and it is defined in Eq.~(\ref{eq:reference}). 
One can easily confirm that $\ket{a,k}$ in Eq.~(\ref{eq:akstate}) is an eigenstate of the translational operator, i.e., 
$\hat{T}\ket{a,k}=e^{ik}\ket{a,k}$. 
If the reference state $\ket{a}$ is not compatible with the momentum $k$, the state $\ket{a,k}$ generated in Eq.~(\ref{eq:akstate}) 
vanishes. 
The compatibility of the chosen reference state $\ket{a}$ and the normalization factor $N^{~}_{a,k}$ can be determined as 
\begin{equation}
N^{~}_{a,k}  =  \frac{ N \left| \bra{a} \sum_{j=1}^{N} e^{-ikj}_{~} \hat{T}^{j}_{~} \ket{a} \right|^2}{\bra{a} \sum_{j=1}^{N} \hat{T}^{j}_{~} \ket{a}}.
\end{equation}
If this quantity is zero, it implies that the chosen reference state $\ket{a}$ is not compatible with the momentum $k$. 
Otherwise, this quantity gives the normalization factor of the state $\ket{a,k}$.

Considering the cyclicity of the translated states, $\{\ket{a_j} \equiv \hat{T}^j\ket{a}\}^{~}_{1\leq j \leq N}$, 
we can simply choose only one state as a representative among $\{\ket{a_j}\}^{~}_{1\leq j \leq N}$. 
The QS$^3$ package chooses one state with the smallest integer $a$, i.e., $\ket{\mathfrak{a}} \equiv \ket{\min_j a_j}$, 
which is used to generate $\ket{a,k}$ in Eq.~(\ref{eq:akstate}). 
We have to check all possible $\combin{N}{\nod}$-states, $\{ \ket{a} \}$, in this way, and determine which states and how many states 
are representatives in the target subspace specified with ($k$, $\nod$). 
The QS$^3$ package uses Algorithm~\ref{algorithm:check_state} to check whether or not a state is representative and evaluate the corresponding normalization factor, and uses Algorithm~\ref{algorithm:mk_list} to make a list of the representative states $\{\ket{\repa} \}$ and a list of the corresponding normalization factors $\{ N_{\repa,k}\}$, representing the symmetry-adapted basis sets $\{\ket{\repa,k}\}$.

\begin{figure}[h]
\begin{algorithm}[H]
  \caption{Check whether a state $|a\rangle$ is representative, and evaluate the corresponding normalization factor.}
  \label{algorithm:check_state}
   \begin{algorithmic}[1]
    \Require{integer $a$ with $1 \leq a \leq \combin{N}{\nod}$, and real $k \in \{2\pi K/N\}^{~}_{0 \leq K < N}$}
    \Ensure{real $N^{~}_{a,k} \geq 0$.}
    \Function{check\_state}{$a,k$}
    \State ${\bm n}=${\sc f\_bar}$(a,\nod,N)$
    \State $N_{a,k}:=0;~c^*:=0;~n_c:=0$
	\For {$j = 1$ to N }
		\State ${\bm n}:=${\sc shift\_func}$({\bm n})$
		\Statex \Comment The function {\sc shift\_func}$({\bm n})$ gives a $\nod$-dimensional vector corresponding to a translated state, $\hat{T} \ket{a}$.		
		\State ${\bm n}:=${\sc insertion\_sort}$({\bm n})$
		\Statex \Comment The function {\sc insertion\_sort}$({\bm n})$ sorts the vector elements in ascending order by using the insertion sort algorithm. When $\nod$ is $\mathcal{O}(1)$, we confirm that the insertion sort algorithm is generally faster than the quick sort algorithm. 
		\State $a' := F({\bm n})$ 
		\If {$a' < a$}
		\State \Return($N^{~}_{a,k}=0$)
		\Else 
		\If {$a'=a$}
		\State $c^*:=c^*+e^{ikj}_{~}$; $n^{~}_c:=n^{~}_c+1$
		\EndIf
		\EndIf
	\EndFor
	\State $N^{~}_{a,k} = |c^*|^2_{~} N / n^{~}_c $
	\State \Return ($N^{~}_{a,k}$)
    \EndFunction    
   \end{algorithmic}
\end{algorithm}
\end{figure}

\begin{figure}[h]
\begin{algorithm}[H]
  \caption{Making lists of representative states and the corresponding normalization factors.}
  \label{algorithm:mk_list}
   \begin{algorithmic}[1]
    \Require{real $k$}
    \Ensure{integer $d$ with $0 \leq d \leq \combin{N}{\nod}$, $d$-dimensional integer vector ${\bm \sigma}$, and $d$-dimensional non-negative real vector ${\bm R}$}
    \Function{mk\_list}{$k$}
      \State $d := 0$
	\For {$ a = 1$ to $\combin{N}{\nod}$}
		\State $N^{~}_{a,k}:=${\sc check\_state}$(a,k)$
		\If {$N^{~}_{a,k}>0$}
		\State $d := d + 1$
		\State $\sigma^{~}_d:=a$
		\State $R^{~}_d:=\sqrt{N^{~}_{a,k}}$
		\EndIf
	\EndFor
	\State \Return($d,{\bm \sigma},{\bm R}$)
    \EndFunction    
   \end{algorithmic}
\end{algorithm}
\end{figure}

We are now ready to construct the Hamiltonian matrix based on the symmetry-adapted basis sets $\{\ket{\repa,k}\}$. 
A state obtained after operating the Hamiltonian $\hat{\mathcal{H}}$ to each basis state $\ket{\repa,k}$ is given by
\begin{equation}
\hat{\mathcal{H}}\ket{\repa,k}=\frac{1}{\sqrt{N^{~}_{\repa,k}}} \sum_{n}^{~} \sum_{j=1}^{N} e^{-ikj}_{~} \hat{T}^{j} \hat{h}_n \ket{\repa}, 
\label{eq:hak}
\end{equation}
where
\begin{equation}
\hat{h}_n \ket{\repa} = \frac{J^{xy}_n}{2} (\hat{s}^+_{r^{~}_n} \hat{s}^-_{r^{\prime}_n} + \hat{s}^-_{r^{~}_n} \hat{s}^+_{r^{\prime}_n})\ket{\repa} + J^{z}_n \hat{s}^z_{r^{~}_n} \hat{s}^z_{r^{\prime}_n}\ket{\repa}. 
\label{eq:hna}
\end{equation}
Note that the Zeeman term in Eq.~(\ref{eq:xxz_hamiltonian}) can be treated separately because the U(1) symmetry is adapted in the basis sets. 
We should also note that the off-diagonal term in Eq.~(\ref{eq:hna}) flips a spin in the representative state $\ket{\repa}$ and the generated state, 
$\ket{a^{(n)}_{~}}=(\hat{s}^+_{r^{~}_n} \hat{s}^-_{r^{\prime}_n} + \hat{s}^-_{r^{~}_n} \hat{s}^+_{r^{\prime}_n})\ket{\repa}$, 
is not necessarily a representative state. 
Therefore, we have to check if the flipped state $\ket{a^{(n)}_{~}}$ is compatible with the momentum $k$. 
If it is the case, we have to seek the representative state $\ket{\repa^{(n)}_{~}} \equiv \ket{\min_j a^{(n)}_{j}}$ by applying translational operations onto $\ket{a^{(n)}_{~}}$, i.e., $\ket{a^{(n)}_{j}} \equiv \hat{T}^j \ket{a^{(n)}_{~}} $. 

Consequently, we can write Eq.~(\ref{eq:hak}) as 
\begin{eqnarray}
\hat{H}\ket{\repa,k} & = & 
\sum_{n}^{~} \frac{J^{xy}_n}{2} e^{-ik\ell^{~}_{n}}_{~} \sqrt{\frac{N^{~}_{\repa^{(n)},k}}{N^{~}_{\repa,k}}} \left( 1-\delta_{f^{~}_{n,\repa},f^{\prime}_{n,\repa}} \right) \ket{\repa^{(n)}_{~},k} \nonumber \\
&&+ \sum_{n}^{~}  \bra{\repa} J^z_n \hat{s}^z_{r^{~}_{n}} \hat{s}^z_{r^{\prime}_{n}} \ket{\repa} ~ \ket{\repa,k}~,
\label{eq:hak2}
\end{eqnarray}
where $\ell_n$ is obtained from the relationship $\ket{\repa^{(n)}_{~}} = T^{\ell^{~}_n}_{~} \ket{a^{(n)}_{~}}$ with $1 \leq \ell^{~}_n \leq N$.
The two variables $f^{~}_{n,\repa}$ and $f^{\prime}_{n,\repa}$ are logical ones given by 
$(p^{~}_{n,\repa},f^{~}_{n,\repa}):=${\sc binary\_search}$(r^{~}_n,${\sc f\_bar}$(\repa,\nod,N),1,\nod)$ and 
$(p^{\prime}_{n,\repa},f^{\prime}_{n,\repa}):=${\sc binary\_search}$(r^{\prime}_n,${\sc f\_bar}$(\repa,\nod,N),1,\nod)$, respectively. 
These variables are used to judge if each off-diagonal term in the Hamiltonian contributes. 
In the QS$^3$ package, Algorithm~\ref{algorithm:representative} is used to search the representative state $\ket{\repa^{(n)}_{~}}$ 
for the off-diagonal matrix elements and to obtain the corresponding $\ell_n$ value, and 
Algorithm~\ref{algorithm:full_hamiltonian_sym_adapt} is to construct the Hamiltonian matrix. 
Note the order of two for-loops associated with $\syma$ and $n$ in Algorithm~\ref{algorithm:full_hamiltonian_sym_adapt} 
that is chosen to enhance the performance of open MP parallelization 
applying to $\syma$.

\begin{figure}[h]
\begin{algorithm}[H]
  \caption{Seeking the representative state $\repa^{(n)}_{~}$ and the $\ell^{~}_n$ value for operations in Eq.~(\ref{eq:hak2})}
  \label{algorithm:representative}
   \begin{algorithmic}[1]
    \Require{integer $a^{(n)}_{~}$ and $\nod$}
    \Ensure{integer $\repa^{(n)}_{~}$ and $\ell^{~}_{n}$.}
    \Function{representative}{$a^{(n)}_{~},\nod $}
    \State $\repa^{(n)}_{~} := a^{(n)}_{~}$
    \State ${\bm n}:=${\sc f\_bar}$(a^{(n)}_{~},\nod,N)$
	\For {$j = 1$ to $N$}
		\State ${\bm n}:=${\sc shift\_func}$({\bm n})$
		\State ${\bm n}:=${\sc insertion\_sort}$(\bm n)$
		\State $a := F({\bm n})$
		\If {$a \leq \repa^{(n)}_{~}$}
		\State $\repa^{(n)}_{~} := a$; $\ell^{~}_{n} := j$
		\EndIf
	\EndFor
	\State \Return($\repa^{(n)},\ell^{~}_{n}$)
    \EndFunction    
   \end{algorithmic}
\end{algorithm}
\end{figure}
\begin{figure}[h]
\begin{algorithm}[H]
  \caption{Generation of Hamiltonian matrix with the symmetry-adapted basis sets.}
  \label{algorithm:full_hamiltonian_sym_adapt}
   \begin{algorithmic}[1]
    \Require{$\noi$-dimensional integer vectors ${\bm r}$ and ${\bm r}'$, $\noi$-dimensional real vectors ${\bm J}^{xy}_{~}$ and ${\bm J}^{z}_{~}$, $d$-dimensional integer vector ${\bm \sigma}$ and real vector ${\bm R}$ for the lists of the representative states and their normalization factors, respectively, and real $k$}
    \Ensure{$d$-dimensional complex matrix ${\bf H}=\{ h_{\syma,\syma'}\}$.}
    \Function{gen\_full\_ham\_sym\_adapt}{${\bm r},{\bm r}',{\bm J}^{xy}_{~},{\bm J}^{z}_{~},{\bm \sigma},{\bm R},k$}
    	\State ${\bf H}:=0$
	\For {$\syma = 1$ to $d$}
		\State ${\bm n}=${\sc f\_bar}$(\sigma_\syma,\nod,N)$
		\For {$n = 1$ to $\noi$}
			\State $(p,f):=${\sc binary\_search}$(r^{~}_n,{\bm n},1,\nod)$
			\State $(p',f'):=${\sc binary\_search}$(r'_n,{\bm n},1,\nod)$
			\If {$f=f'$}
				\State $h^{~}_{\syma,\syma} := h^{~}_{\syma,\syma}  + J^z_n/4$
			\Else
				\State $h^{~}_{\syma,\syma} := h^{~}_{\syma,\syma} - J^z_n/4$
				\State $a':=F$({\sc spin\_exchange}$(r,p,f,r',p',{\bm n}))$
				\State $(\repa',\ell):=${\sc representative}$(a',\nod)$	
				\State $(\syma',f):=${\sc binary\_search}$(\repa',{\bm \sigma},1,d)$
				\If {$f={\rm True}$}
				\State $h^{~}_{\syma,\syma'} := h^{~}_{\syma,\syma'} + \frac{J^{xy}_n}{2} e^{ik\ell}\sqrt{R^{~}_{\syma'}/R^{~}_{\syma}}$ 
				\EndIf
			\EndIf
		\EndFor
	\EndFor
	\State \Return{(${\bf H}$)}
    \EndFunction
   \end{algorithmic}
\end{algorithm}
\end{figure}

\subsection{Full diagonalization}
One can full diagonalize the whole Hamiltonian 
to obtain all eigenvalues $\{E_{\nu} \}$ 
and the corresponding eigenvectors $\{ \ket{\nu} \}$ 
by separately diagonalizing block diagonalized Hamiltonian matrices constructed via  
Algorithm~\ref{algorithm:full_hamiltonian} or Algorithm~\ref{algorithm:full_hamiltonian_sym_adapt} with different symmetry sectors. 
Accordingly, one can for example compute the temperature dependence of any physical quantity $\hat{A}$ based on the thermal average,
\begin{equation}
\langle \hat{A} \rangle^{\rm ens}_{\beta, N} = \sum_{\nu} \frac{ e^{-\beta E_{\nu}} }{Z(\beta)} \langle \nu | \hat{A} | \nu \rangle,
\label{eq:ens}
\end{equation}
where $\beta$ is the inverse temperature, $Z(\beta)=\sum_\nu e^{-\beta E_{\nu}}$ is the partition function, and the summation of $\nu$ 
runs over all symmetry sectors with different values of $\nod$ and/or $k$. 
However, note that the accessible matrix dimension is very limited in the full diagonalization calculation, typically up to $O(10^4)$ on a currently available standard computer. 
Therefore, one may not be able to treat all subspaces of the Hamiltonian even when the Hamiltonian matrix is 
block diagonalized with different symmetry sectors. 

The QS$^3$ package is specialized for the system under a high magnet field, in which one can treat much larger system sizes 
near the saturation field. This implies that one may access the finite temperature physics of large systems but at sufficiently low 
temperature where the low-energy eigenvalues are reasonably separated from those for the symmetry sectors with larger 
$\nod$ 
and thus the latter contribution to the thermal average $\langle \hat{A} \rangle^{\rm ens}_{\beta, N}$ 
can be simply discarded.

The QS$^3$ package uses DHEEVR/ZHEEVR routine in LAPACK~\cite{anderson1999} for the full 
diagonalization to obtain all eigenvalues and eigenvectors of the Hamiltonian matrix constructed with the 
symmetry-adapted basis sets.

\subsection{Multiplying Hamiltonian to state vectors (matrix-vector product)}
In order to calculate the lowest eigenvalue (and also the several lowest eigenvalues) and the corresponding eigenvector(s) 
of the Hamiltonian matrix, one can also employ the conventional Lanczos method, instead of the full diagonalization, which allows us to 
treat larger system sizes. 
The main and most time-consuming part in the Lanczos method is a matrix-vector product, 
and the QS$^3$ package does this operation based on the symmetry-adapted basis sets. 
Assuming that both the U(1) and translational symmetries are adapted, the resulting vector after the matrix-vector product operation 
$ \hat{H} \ket{\phi}$ is expressed with the basis sets $\{\ket{\repa,k} \}$ and each element $\psi^{~}_{\repa,k}$ can be obtained 
as 
\begin{eqnarray}
\psi^{~}_{\repa,k} & =& \braket{\repa,k}{\psi} = \bra{\repa,k} \hat{H} \ket{\phi} \nonumber \\
& = & 
\sum_{n}^{~} \frac{J^{xy}_n}{2} e^{ik\ell^{~}_{n}}_{~}
 \sqrt{\frac{N^{~}_{\repa^{(n)}_{~},k}}{N^{~}_{\repa,k}}}  
\left( 1-\delta_{f^{~}_{n,\repa},f^{\prime}_{n,\repa}} \right) \phi_{\repa^{(n)}_{~},k}^{~}  \nonumber \\
&&+ \sum_{n}^{~}  \bra{\repa} J^z_n \hat{s}^z_{r^{~}_{n}} \hat{s}^z_{r^{\prime}_{n}} \ket{\repa} ~ \phi_{\repa,k}^{~} ~,
\end{eqnarray}
where a state vector $\ket{\phi}=\sum_\repa \phi_{\repa,k} \ket{\repa,k}$ is an input vector. 
The QS$^3$ package uses Algorithm~\ref{algorithm:hphi} to do this procedure. 
Note that the calculation of each element is done on the fly and hence the accessible vector dimension can be enlarged up to 
$O(10^8)$.

\begin{figure}[h]
\begin{algorithm}[H]
  \caption{Perform ${\bm \psi}:={\bf H}{\bm \phi}$}
  \label{algorithm:hphi}
   \begin{algorithmic}[1]
    \Require{integer vectors ${\bm \sigma}$, ${\bm r}$, and ${\bm r}'$, real $k$, real vectors ${\bm R}$, ${\bm J}^{xy}_{~}$, and ${\bm J}^{z}_{~}$ as in Algorithm~\ref{algorithm:full_hamiltonian_sym_adapt}, and $d$-dimensional complex vector ${\bm \phi}$.}
    \Ensure{$d$-dimensional complex vector ${\bm \psi}$.}
    \Function{ham\_to\_vec}{${\bm r},{\bm r}',{\bm J}^{xy}_{~},{\bm J}^{z}_{~},{\bm \sigma},k,{\bm R},{\bm \phi}$}
    	\State ${\bm \psi}:=0$
	\For {$\syma = 1$ to $d$}
		\State ${\bm n}=${\sc f\_bar}$(\sigma_\syma,\nod,N)$
		\For {$n = 1$ to $\noi$}
			\State $(p,f):=${\sc binary\_search}$(r^{~}_n,{\bm n},1,\nod)$
			\State $(p',f'):=${\sc binary\_search}$(r'_n,{\bm n},1,\nod)$
			\If {$f=f'$}
				\State $\psi^{~}_{\syma} := \psi^{~}_{\syma} + \frac{J^z_n}{4} \phi^{~}_{\syma}$
			\Else
				\State $\psi^{~}_{\syma} := \psi^{~}_{\syma} - \frac{J^z_n}{4} \phi^{~}_{\syma}$
				\State $a':=F$({\sc spin\_exchange}$(r,p,f,r',p',{\bm n}))$
				\State $(\repa',\ell):=${\sc representative}$(a',\nod)$	
				\State $(\syma',f):=${\sc binary\_search}$(\repa',{\bm \sigma},1,d)$
				\If {$f={\rm True}$}
				\State $\psi^{~}_\syma := \psi^{~}_\syma + \frac{J^{xy}_n}{2} e^{ik\ell} \sqrt{R^{~}_{\syma'}/R^{~}_{\syma}} \phi^{~}_{\syma'}$ 
				\EndIf
			\EndIf
		\EndFor
	\EndFor
	\State \Return(${\bm \psi}$)
    \EndFunction
   \end{algorithmic}
\end{algorithm}
\end{figure}

\subsection{Calculating expectation values}
The QS$^3$ package can evaluate the local magneitzation $\bra{\phi} \hat{s}^z_{{r}^{~}} \ket{\phi}$ and the two-point spin correlation 
function $\bra{\phi} \hat{s}^\alpha_{{r}^{~}} \hat{s}^\beta_{{r'}^{~}} \ket{\phi}$ where $(\alpha,\beta)\in\{(z,z), (\pm,\mp)\}$ 
after computing eigenvectors $\ket{\phi}$ of the Hamiltonian matrix. 
When the eigenvector $\ket{\phi}=\sum_{\repa} \phi^{~}_{\repa,k} \ket{\repa,k}$ respects the translational symmetry 
with the momentum $k$, 
the expectation value of a operator preserving the translational symmetry can be evaluated simply by reusing 
Algorithm~\ref{algorithm:full_hamiltonian_sym_adapt}, where the matrix elements of the Hamiltonian matrix 
are evaluated in the symmetry-adapted basis sets.
Therefore, in the QS$^3$ package, the translationally-symmetrized operators 
$\frac{1}{N}\sum_{j=1}^{N} \hat{T}^{j}_{~} \hat{O} \hat{T}^{-j}_{~}$ with $\hat{O}=\hat{s}^z_{{r}^{~}}$ and 
$\hat{s}^\alpha_{{r}^{~}} \hat{s}^\beta_{{r'}^{~}}$ are used, instead of directly treating the local operators $\hat{O}$, 
for the expectation values: 
$\bra{\phi} \hat{O} \ket{\phi} = \frac{1}{N}  \bra{\phi} \left( \sum_{j=1}^{N} \hat{T}^{j}_{~} \hat{O} \hat{T}^{-j}_{~} \right) \ket{\phi}$.

\subsection{Thick-restart Lanczos method}
The QS$^3$ package employs the thick-restart Lanczos method~\cite{Wu1999,Wu2000} to compute the multiple lowest eigenvalues and the corresponding eigenvectors of the Hamiltonian matrix, i.e., the ground state and the several lowest-excited states of the Hamiltonian $\hat{H}$.  
The algorithm is provided in Algorithm~\ref{algorithm:TRL}.
In the first part of this algorithm, exactly the same procedure of the conventional Lanczos method is employed to 
generate $N^{~}_{M}+1$ Lanczos vectors, ${\bf \Psi}=\{\bm \psi_x\}^{~}_{1 \leq x \leq N^{~}_{M}}$ and ${\bm \psi}^{~}_{N^{~}_{M}+1}$, 
and construct the tridiagonal matrix
\begin{equation}
{\bf T} = {\bf \Psi}^{\dagger} {\bf H} {\bf \Psi} := 
\begin{pmatrix}
\alpha_1 & \beta_1  &                               \\
\beta_1^*  & \alpha_2 & \beta_2 &               \\
              & \ddots     & \ddots   & \ddots & \\
              &                & \beta^{*}_{N^{~}_{\rm M}-2}  & \alpha^{~}_{N^{~}_{\rm M}-1} & \beta^{~}_{N^{~}_{\rm M}-1} \\
               &             &                & \beta^{*}_{N^{~}_{\rm M}-1}  & \alpha^{~}_{N^{~}_{\rm M}}  \\
\end{pmatrix}~,
\end{equation}
where $\alpha_x = {\bm \psi}^{\dagger}_x {\bf H} {\bm \psi}^{~}_x$, $\beta_x = {\bm \psi}^{\dagger}_x {\bf H} {\bm \psi}^{~}_{x+1}$, 
and other elements are zero. 
Then, this tridiagonal matrix ${\bf T}$ is diagonalized to obtain the eigenvalues ${\bm e}=\{ e_x \}$ in ascending order and 
the unitary matrix ${\bf C}=\{c^{~}_{x,x'}\}$ such that ${\bf T}={\bf C} \big( {\rm diag}[{\bm e}] \big) {\bf C}^{\dagger}_{~}$. 

The thick-restart Lanczos method focuses on the lowest $N^{~}_{\rm K}\,(<N^{~}_{\rm M})$ eigenvalues 
and the corresponding eigenvectors by the keeping $N^{~}_{\rm K}+1$ vectors, 
$\{ \{{\bm \psi}_y\}^{~}_{1 \leq y \leq N^{~}_{K}},{\bm \psi}^{~}_{N^{~}_{K}+1} \}:=\{ {\bf \Psi}\{c^{~}_{x,y}\}_{1 \leq y \leq N^{~}_{K}}, {\bm \psi}^{~}_{N^{~}_{M}+1}\}$,  
and generates the $N^{~}_{M} - N^{~}_{K}$ Lanczos vectors $\{{\bm \psi}^{~}_{N^{~}_{K}+2},\cdots, {\bm \psi}^{~}_{N^{~}_{M}+1}\}$, 
according to the procedures in the conventional Lanczos method with the initial vector ${\bm \psi}^{~}_{N^{~}_{M}+1}$, 
and also the matrix ${\bf T}$ as 
\begin{equation}
{\bf T} := \begin{pmatrix}
e_1 &            &                                      & \beta_1 \\
       & \ddots &                                     & \vdots    \\
       &            & e^{~}_ {N^{~}_{K}}  & \beta^{~}_{N^{~}_{K}} & \\
\beta^*_1 & \hdots  & \beta^{*}_{N^{~}_{\rm K}}  & \alpha^{~}_{N^{~}_{\rm K}+1} & \beta^{~}_{N^{~}_{\rm K}+1} &  \\
 &   &   &  \ddots & \ddots & \ddots  \\
 &&              &                & \beta^{*}_{N^{~}_{\rm M}-2}  & \alpha^{~}_{N^{~}_{\rm M}-1} & \beta^{~}_{N^{~}_{\rm M}-1} \\
 &&             &             &                & \beta^{*}_{N^{~}_{\rm M}-1}  & \alpha^{~}_{N^{~}_{\rm M}}
\end{pmatrix}~,
\end{equation}
where $\{ \beta^{~}_y \}_{1 \leq y \leq N^{~}_{\rm K}}:=\{ \beta^{~}_{N^{~}_{\rm M}} c^{~}_{N^{~}_{\rm M}, y}\}$ and only elements 
not generally zero are shown. 
This matrix ${\bf T}$ is then diagonalized to obtain the eigenvalues ${\bm e}=\{ e_x \}$ in ascending order 
and the corresponding unitary matrix ${\bf C}=\{c^{~}_{x,x'}\}$.  
This procedure is repeated until the lowest $N^{~}_{\rm K}$ eigenvalues are converged within the specified convergence ratio 
$\epsilon$ or the total number of iterations exceeds a given integer $I^{~}_{\rm M}$. 
This is the second part of the algorithm described in Algorithm~\ref{algorithm:TRL}. 
Therefore, the thick-restart Lanczos method requires maximally the $N^{~}_{\rm M}(>N^{~}_{\rm K})$ dimensional Krylov subspace. 
The integer numbers $N^{~}_{\rm M}$ and $I^{~}_{\rm M}$ and the real number $\epsilon$ are input parameters, which determine 
the quality of the calculation.

\begin{figure}[h]
\begin{algorithm}[H]
  \caption{Thick-restart Lanczos method}
  \label{algorithm:TRL}
   \begin{algorithmic}[1]
    \Require{integer $N_{\rm K}\,(>0)$, $N_{\rm M}\,( > N_{\rm K})$, $d$, and $I_{\rm M}\,(>0)$, and real $\epsilon \ll 1$}
    \Ensure{real ${\bm e}=\{e^{~}_x\}_{1 \leq x \leq N_{\rm M}}$ and complex ${\bf \Psi} = \{ \psi_{ax} \}^{1 \leq a \leq d}_{1 \leq x \leq N_{\rm M}}$.}
    \Function{thick\_restart\_lanczos}{$N_{\rm K}, N_{\rm M}, I_{\rm M}, \epsilon$}
    	\State ${\bm \psi}_x=\{\psi_{ax}\}^{~}_{1\leq a \leq d}$; ${\bm \psi}_1:=${\sc random\_vec}$(d)$
    	\Statex \Comment The function {\sc random\_vec}$(d)$ returns $d$-dimensional complex random vector.
    	\State $\beta_0:=\sqrt{|{\bm \psi}_{1}|^2}$
	\For {$x = 1$ to $N_{\rm M}$}
		\State ${\bm \psi}^{~}_x:={\bm \psi}^{~}_x/\beta^{~}_{x-1}$
		\State ${\bm v}^{~}:=${\sc ham\_to\_vec}(${\bm r},{\bm r}',{\bm J}^{xy}_{~},{\bm J}^{z}_{~},{\bm \sigma},k,{\bm \psi}_{x}$)
		\State $\alpha^{~}_x:={\bm \psi}_x^\dagger \cdot {\bm v}$
		\State ${\bm \psi}^{~}_{x+1}:=\left\{ \begin{matrix}
		{\bm v}^{~} - \alpha^{~}_x {\bm \psi}^{~}_x & (x=1) \\
		{\bm v}^{~} - \alpha^{~}_x {\bm \psi}^{~}_x - \beta^{~}_{x-1} {\bm \psi}^{~}_{x-1}& (x>1) \\
		\end{matrix}\right.$
		\State ${\bm \psi}^{~}_{x+1} :=${\sc reorthogonalization}$({\bm \Psi},x+1)$
		\Statex \Comment The function {\sc reorthogonalization}$({\bm \Psi},x)$ performs reorthogonalization, for example, with the modified Gram-Schmidt procedure, to numerically keep the orthogonality ${\bm \psi}_{x'<x}^{\dagger} \cdot {\bm \psi}^{~}_{x} =0$. 
		\State $\beta^{~}_x:=\sqrt{|{\bm \psi}^{~}_{x+1}|^2}$
	\EndFor
	\State ${\bm \psi}^{~}_{N^{~}_{{\rm M}+1}}:={\bm \psi}^{~}_{N^{~}_{{\rm M}+1}}/\beta^{~}_{N^{~}_{{\rm M}}}$
	\State $({\bm e}, {\bm C}=\{c_{x,x'}\}):=${\sc diag\_tri}$(\{\alpha^{~}_x \},\{\beta^{~}_x\})$
    	\Statex \Comment The function {\sc diag\_tri}$(\{\alpha^{~}_x \},\{\beta^{~}_x\})$ returns eigenvalues ${\bm e}$ (ascending order) and the corresponding eigenvectors ${\bm C}$ of a real symmetric tridiagonal matrix with diagonal elements $\{\alpha^{~}_x \}$ and sub-diagonal elements $\{\beta^{~}_x\}$.
	\For {$I = 1$ to $I_{\rm M}$}
		\State $\{\bm \psi^{~}_y\}^{~}_{1 \leq y \leq N^{~}_{\rm K}} := {\bf \Psi}\{c^{~}_{x,y}\}$ 
		\If {$I=1$}
		        \State ${\bm e'}=\{ e^{~}_y \}^{~}_{1 \leq y \leq N^{~}_{\rm K}}$
		\Else
		\If {$\max_{1 \leq y \leq N^{~}_{\rm K}} | e'_y/e^{~}_y - 1 | < \epsilon$ }
			\State Exit
		\Else
		         \State ${\bm e'}=\{ e^{~}_y \}^{~}_{1 \leq y \leq N^{~}_{\rm K}}$
		\EndIf
		\EndIf  
		\State $\{\alpha_y\}^{~}_{1 \leq y \leq N^{~}_{\rm K}}:=\{ e^{~}_y \}$
		\State ${\bm \psi}^{~}_{N^{~}_{\rm K}+1}:={\bm \psi}^{~}_{N^{~}_{\rm M}+1}$
		\State $\{ \beta^{~}_y \}_{1 \leq y \leq N^{~}_{\rm K}}:=\{ \beta^{~}_{N^{~}_{\rm M}} c^{~}_{N^{~}_{\rm M}, y}\}$
		\State lines 6 and 7 with $x=N^{~}_{\rm K}+1$.
		\State ${\bm v} := {\bm v}^{~} -\sum_{y}\beta_y {\bm \psi}_y$
		\State ${\bm \psi}^{~}_{N^{~}_{\rm K}+2}:= {\bm v} - \alpha^{~}_{N^{~}_{\rm K}+1} {\bm \psi}^{~}_{N^{~}_{\rm K}+1}$
		\State ${\bm \psi}^{~}_{N^{~}_{\rm K}+2} :=${\sc reorthogonalization}$({\bm \Psi},N^{~}_{\rm K}+2)$
		\State $\beta^{~}_{N^{~}_{\rm K}+1}:=\sqrt{|{\bm \psi}^{~}_{N^{~}_{\rm K}+2}|^2}$
		\State lines 4-12 with the starting value of $x=N^{~}_{\rm K}+2$.
		\State ${\bf T}:=0$
		\State $\{ t^{~}_{xx} \}_{1 \leq x \leq N^{~}_{\rm M}}=\{\alpha^{~}_x\}$
		\State $\{ t^{~}_{y,N^{~}_{\rm K}+1} \}_{1 \leq y \leq N^{~}_{\rm K}} = \{ t^*_{N^{~}_{\rm K}+1,y} \} := \{ \beta^{~}_y \}$
		\State $\{ t_{z,z+1} \}_{N^{~}_{\rm K}+1 \leq z \leq N^{~}_{\rm M}-1} = \{ t^*_{z+1,z} \} := \{ \beta_z \}$
		\State $({\bm e}, {\bm C}):=${\sc diag}$({\bf T})$
	    	\Statex \Comment The function {\sc diag}$({\bf T})$ returns eigenvalues (ascending order) and the corresponding eigenvectors of Hermitian matrix ${\bf T}$.
    \EndFor    
	\State \Return(${\bm e},{\bm \Psi}$)
\EndFunction
\end{algorithmic}
\end{algorithm}
\end{figure}

\subsection{Multiplying an operator to state vectors represented with the symmetry-adapted basis sets}
The QS$^3$ package computes the static and dynamical spin structure factors after obtaining a target eigenvector $\ket{\phi}$ 
of the Hamiltonian matrix. Considering a periodic chain with $N$ spins, as an example, 
the Fourier transform of the spin operator at wave number $q$ with $\alpha=\pm,\, z$ is given as 
\begin{equation}
\hat{S}^{\alpha}_{q} = \frac{1}{\sqrt{N}} \sum_{j=1}^N e^{-iqj}_{~} \hat{T}^{j}_{~} \hat{s}^{\alpha}_1 \left( \hat{T}^{j}_{~} \right)^\dagger_{~}.
\end{equation}
One can easily show that the operator $\hat{S}^{\alpha}_{q}$ satisfies the following relation: 
\begin{equation}
\hat{S}^{\alpha}_{q} \hat{T}^{j} = e^{-iqj} \hat{T}^{j}_{~} \hat{S}^{\alpha}_{q}.  
\end{equation}
Using this relation, one of the basic operations, $\hat{S}^{\alpha}_{q} \ket{\phi}$, necessary for computing the static and dynamical spin structure factors can be rewritten as 
\begin{equation}
\hat{S}^{\alpha}_{q} \ket{\phi}  =  \sum_{\repa} \phi^{~}_{\repa,k} \hat{S}^{\alpha}_{q} \ket{\repa,k}
\end{equation}
with 
\begin{eqnarray}
\hat{S}^{\alpha}_{q} \ket{\repa,k} & = & 
\frac{1}{\sqrt{N^{~}_{\repa,k}}} \sum_{j}^{~} e^{-ikj}_{~} \hat{S}^{\alpha}_{q} \hat{T}^{j} \ket{\repa} \nonumber \\
& = & \frac{1}{\sqrt{N^{~}_{\repa,k}}} \sum_{j}^{~} e^{-i(k+q)j}_{~} \hat{T}^{j} \hat{S}^{\alpha}_{q} \ket{\repa}  \nonumber \\
& = & \frac{1}{\sqrt{N^{~}_{\repa,k}N}} \sum_{j,j'}^{~} e^{-i(k+q)j}_{~} \hat{T}^{j} e^{-iqj'} \hat{T}^{j'}_{~} \hat{s}^{\alpha}_1 \left(\hat{T}^{j'}_{~}\right)^\dagger_{~} \ket{\repa}, \nonumber \\
\label{eq:sq_phi}
\end{eqnarray}
where $ \ket{\phi}  =  \sum_{\repa} \phi^{~}_{\repa,k} \ket{\repa,k}$ is an eigenvector of the Hamiltonian matrix 
and it is in the subspace specified with momentum $k$ and the number $\nod$ of down spins. 
Note that in computing the transverse components of the spin structure factors with $\alpha=\pm$ in Eq.~(\ref{eq:sq_phi}), 
we have to consider a transition between states with different $U(1)$ symmetry sectors, i.e., from a state $\ket{\repa}$ 
with $\nod$ down spins to a state $\hat{S}^{\alpha=\pm}_{q} \ket{\repa}$ with $\nod\mp1$ down spins.

Let us now introduce the spin state $\ket{b}$ defined as 
\begin{equation}
\hat{T}^{j'}_{~} \hat{s}^{\alpha}_1 \left(\hat{T}^{j'}_{~}\right)^\dagger_{~} \ket{\repa} = c \ket{b}
\end{equation}
with $c=\bra{b} \hat{T}^{j'} \hat{s}^{\alpha}_1 \left(\hat{T}^{j'}_{~}\right)^\dagger_{~} \ket{\repa}$.  
Note that $\ket{b}$ as well as $c$ depends on $\alpha$, $j'$, and $\repa$. 
In general, the state $\ket{b}$ is not the representative state $\ket{\repb} \equiv \ket{\min_j b_{j}}$ for states $\left\{ \ket{b_{j}} \equiv \hat{T}^j \ket{b} \right\}^{~}_{1\leq j \leq N}$, and $\ket{b}$ can be translated to $\ket{\repb}$ by repeatedly applying the translational operator $\hat T$, i.e., $\ket{\repb} = \hat{T}^{\ell} \ket{b}$ with $1 \leq \ell \leq N$, where $\ell$ depends on $\ket{b}$. 
Therefore, we can rewrite Eq.~(\ref{eq:sq_phi}) using the representative state $\ket{\repb}$ as
\begin{eqnarray}
\hat{S}^{\alpha}_{q} \ket{\repa,k}
& = & \frac{1}{\sqrt{N^{~}_{\repa,k}N}} \sum_{j,j'}^{~} e^{-i(k+q)j}_{~} \hat{T}^{j}_{~} e^{-iqj'}_{~} c \hat{T}^{-\ell}_{~} \ket{\repb}. \nonumber \\
& = & \frac{1}{\sqrt{N^{~}_{\repa,k}N}} \sum_{j,j'}^{~} e^{-i(qj'+(k+q)\ell)}_{~} c e^{-i(k+q)(j-\ell)}_{~} \hat{T}^{j-\ell}_{~} \ket{\repb}. \nonumber \\
& = & \frac{1}{\sqrt{N^{~}_{\repa,k}N}} \sum_{j'}^{~} \sqrt{N^{~}_{\repb,k+q}} e^{-i(qj'+(k+q)\ell)}_{~} c \ket{\repb,k+q}. \nonumber \\
\end{eqnarray}
A concrete procedure for performing $\hat{S}^-_{q} \ket{\phi}$ is shown in Algorithm~\ref{algorithm:sq_phi}. 
In the same manner, we can perform $\hat{S}^+_{q}\ket{\phi}$ and $\hat{S}^z_{q}\ket{\phi}$.

\begin{figure}[h]
\begin{algorithm}[H]
  \caption{Perform $\ket{\psi'} := \hat{S}^-_{q} \ket{\phi}$}
  \label{algorithm:sq_phi}
   \begin{algorithmic}[1]
    \Require{real $k$ and $q \in \{2\pi K/N\}^{~}_{0 \leq K < N}$, $d$-dimensional integer vector ${\bm \sigma}$ and real vector ${\bm R}$ for the lists of representative states specified with $\nod$ down spins and their normalization factors, respectively, $d$-dimensional complex vector ${\bm \phi}$, and $d'$-dimensional integer vector ${\bm \sigma'}$ and real vector ${\bm R}'$ for the lists of representative states specified with $\nod+1$ down spins and their normalization factors, respectively.}
    \Ensure{$d'$-dimensional complex vector ${\bm \psi}'$.}
    \Function{smq\_to\_vec}{${\bm \sigma},{\bm R},k,{\bm \phi},{\bm \sigma}',{\bm R}',q$}
    	\State ${\bm \psi}':=0$
	\For {$r = 1$ to $N$}
		\For {$\syma = 1$ to $d$}
			\State ${\bm n}=${\sc f\_bar}$(\sigma_\syma,\nod,N)$
			\State $(p,f):=${\sc binary\_search}$(r,{\bm n},1,\nod)$
			\If {$f={\rm False}$}
				\State ${\bm n}':=(n_1,\cdots,n_{p},r,n_{p+1},\cdots,n_{\nod})$
				\State $b:=F({\bm n}')$
				\State $(\repb,\ell):=${\sc representative}$(b,\nodd)$	
				\State $(\symb,f):=${\sc binary\_search}$(\repb,{\bm \sigma}',1,d')$
				\If {$f={\rm True}$}
				\State $\psi^{\prime}_\symb := \psi^{\prime}_\symb + \sqrt{ \frac{R'_{\symb}}{R^{~}_{\syma}N} } e^{-i(qr+(k+q)\ell)} \phi^{~}_{\syma}$ 
				\EndIf
			\EndIf
		\EndFor
	\EndFor
	\State \Return(${\bm \psi}'$)
    \EndFunction
   \end{algorithmic}
\end{algorithm}
\end{figure}

\subsection{Continued fraction expansion based on the Lanczos algorithm}
\label{sec:continued_fraction_expansion}
Using the continued fraction expansion based on the Lanczos algorithm~\cite{Dagotto94,Haydock72,Gagliano87}, the QS$^3$ package computes the dynamical spin structure factor 
\begin{equation}
S^{\alpha}_{q}(\omega) = -\frac{1}{\pi} {\rm Im}\bra{\phi} \hat{S}^{\alpha \dagger}_{q} \frac{1}{\omega - \hat{H} + E_0 + i\eta } \hat{S}^{\alpha}_{q} \ket{\phi},  
\end{equation}
where $E^{~}_0$ is the ground state energy (i.e., lowest eigenvalue) with the corresponding ground state $\ket{\phi}$ of the Hamiltonian $\hat H$ and positive real number $\eta$ is the broadening factor. 
We can rewrite the above equation as 
\begin{equation}
S^{\alpha}_{q}(\omega) = -\frac{1}{\pi} {\rm Im}
\cfrac{\bra{\phi} \hat{S}^{\alpha\dagger}_{q} \hat{S}^{\alpha}_{q} \ket{\phi}}
{z-\alpha^{~}_1-\cfrac{\beta^2_1}
{z-\alpha^{~}_2-\cfrac{\beta^2_2}
{z-\alpha^{~}_3-\cdots
}}}
\label{eq:cfe}
\end{equation}
with $z=\omega - E^{~}_0 + i\eta$. 
${\bm \alpha}$ and ${\bm \beta}$ in Eq.~(\ref{eq:cfe}) are obtained by the tridiagonalization procedure of the Hamiltonian matrix 
in the Lanczos iteration shown in lines 3-11 of Algorithm~\ref{algorithm:TRL} with the initial state 
$\ket{\psi}^{~}_1=\hat{S}^{\alpha}_{q} \ket{\phi}$ that can be prepared by the procedure shown in Algorithm~\ref{algorithm:sq_phi}. 

\afterpage{\clearpage}

\section{Benchmark results}
\label{sec:benchmark}

\subsection{Parallelization efficiency with openMP}
\begin{figure}[h]
 \begin{center}
  \includegraphics[width=8.0cm]{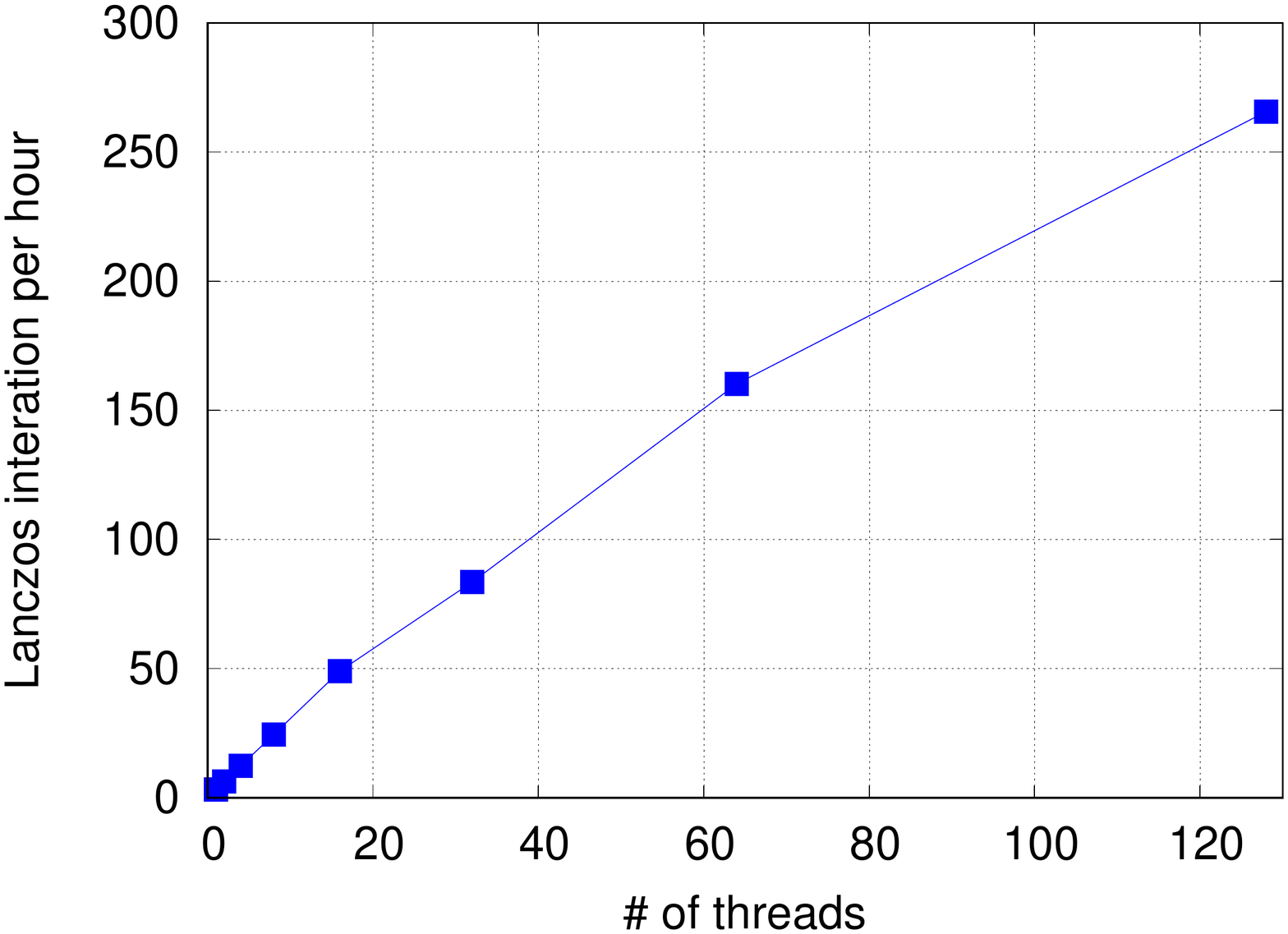}
  \caption{Parallelization efficiency of the QS$^3$ package. The conventional Lanczos algorithm is used to calculated the ground state of an $S=1/2$ isotropic antiferromagnetic Heisenberg model on a simple cubic lattice of 216 sites with $(K_x,K_y,K_z)=(0,0,0)$ and $\nod=5$.}
  \label{fig:Bench}
 \end{center}
\end{figure}

Here we show a benchmark result of the QS$^3$ package for the numerical diagonalization. 
The most time consuming part in the Lanczos algorithm is the Hamiltonian-vector multiplication in Algorithm~\ref{algorithm:hphi} with computational complexity $O( \combin{N}{\nod} \noi \nod \ln \nod )$. 
The QS$^3$ package adopts OpenMP to parallelize this procedure. 
For a typical benchmark, we consider an $S=1/2$ isotropic antiferromagnetic Heisenberg model on a simple cubic lattice of 216 sites ($L_x=L_y=L_z=6$) and calculate the ground state in the subspace with momentum ${\bf k}=(0,0,0)$ and $\nod=5$, by setting the parameters $(K_x, K_y, K_z)=(0, 0, 0)$, using the conventional Lanczos algorithm. 
The dimension of the Hilbert space (i.,e, the Hamiltonian matrix) is $\combin{216}{5}$~(=3,739,729,608) with only adapting the $U(1)$ symmetry and can be reduced 1/$N$ times smaller down to 17,313,563 when the translational symmetry is also adapted. 
Figure~\ref{fig:Bench} shows the efficiency of the parallelization of the QS$^3$ package executed using the supercomputer (Ohtaka) in ISSP with AMD Epyc 7702 2.0~GHz. 
We confirm almost linear acceleration with increasing the number of threads up to 128, although the slope becomes somewhat 
smaller when the number of threads exceeds around 20.

\subsection{Energy-dispersion relation}
One of the essential physical quantities to understand the low-energy physics of a quantum spin model is 
the energy-dispersion relation $E_0({\bm k})$, the ground state energy at each momentum $\bm k$. 
Most of the currently available exact diagonalization libraries compute this quantity but are sufficient for practical use only 
in one-dimensional systems because of the severe limitations of the accessible system sizes. 
The QS$^3$ package can evaluate the energy-dispersion relation around the saturation field even in three-dimensional systems.

\begin{figure}[h]
 \begin{center}
  \includegraphics[width=8.0cm]{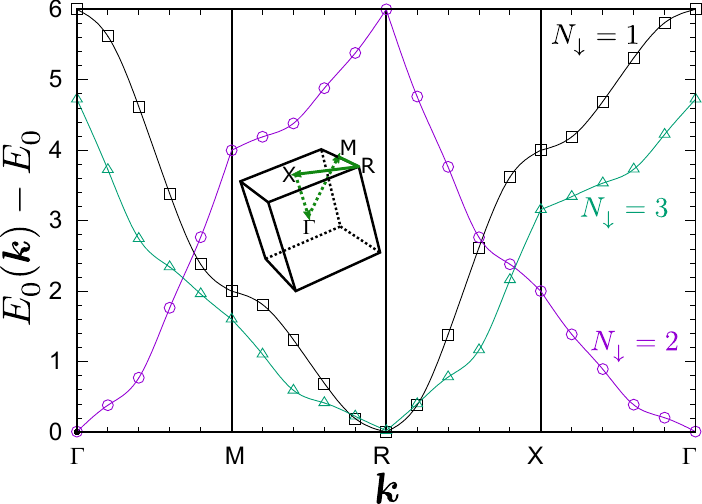}
  \caption{
  Energy-dispersion relation $E_0({\bm k})$ of an $S=1/2$ isotropic antiferromagnetic Heisenberg model on a simple cubic lattice 
  of 1000 sites ($L_x=L_y=L_z=10$) with $\nod$ down spins along the high symmetric momentum ${\bm k}$ line indicated 
  in the inset. The lowest eigenenergies $E_0({\bm k})$ with $\nod=1,2$, and $3$ near the saturation field 
  are plotted relative to the ground state energy $E_0$ of the fully polarized state with $\nod=0$. 
The high symmetric momentum points are indicated by $\Gamma$:~$(0,0,0)$, M:~$(0,\pi,\pi)$, R:~$(\pi,\pi,\pi)$, and X:~$(0,0,\pi)$. 
Solid lines are the cubic-spline interpolation in each path, i.e., $\Gamma \rightarrow {\rm M}$, ${\rm M} \rightarrow {\rm R}$, ${\rm R} \rightarrow {\rm X}$, and  ${\rm X} \rightarrow \Gamma$.
}
  \label{fig:energy_dispersion}
 \end{center}
\end{figure}

For demonstration, we show the energy-dispersion relation for an $S=1/2$ isotropic antiferromagnetic Heisenberg model 
on a simple cubic lattice of 1000 sites ($L_x=L_r=L_z=10$) in Fig.~\ref{fig:energy_dispersion}. 
When only the U(1) symmetry is used, the dimension of the Hilbert space with $\nod=3$ 
is $\combin{1000}{3}=166,167,000$ and approximately 2.5~GByte of physical memory is required to store a state vector 
with the double-complex precision. This implies that the total physical memory up to about 8~GByte is required to obtain 
the ground state by means of the conventional Lanczos method. 
This is rather expensive to perform on a standard laptop computer. 
However, by incorporating the translational symmetry, the required storage per a state vector is reduced down to around 2.5~MByte, 
and thus the computation can be executed easily with a laptop computer.

\subsection{Static and dynamical structure factors}
\begin{figure}[h]
 \begin{center}
  \includegraphics[width=7.5cm]{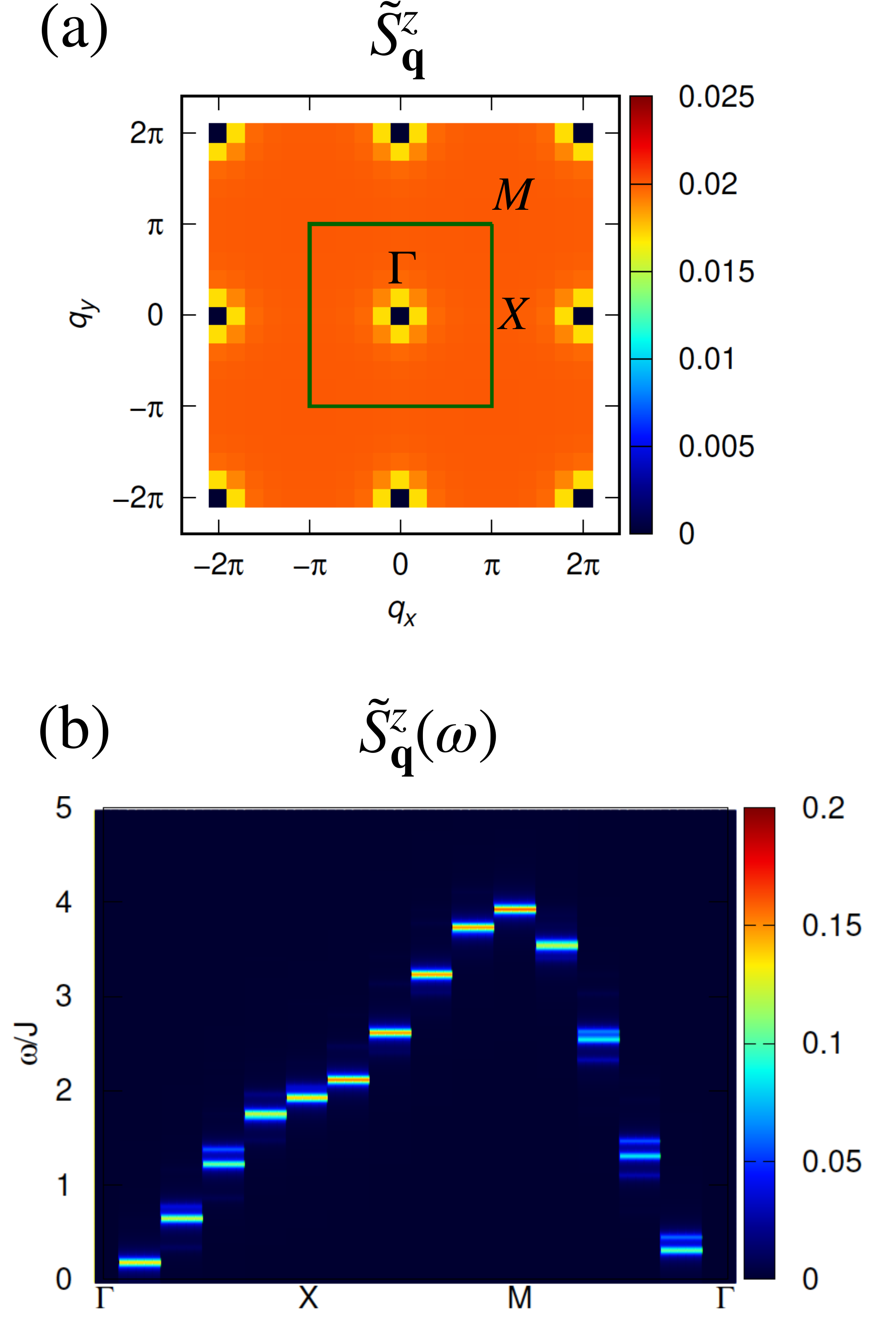}
  \caption{
  (a) The $z$-component of the static spin structure factor $\tilde{S}_{\bf q}^z$ and 
  (b) the $z$-component of the dynamical spin structure factor $\tilde{S}_{\bf q}^z(\omega)$ along the high symmetric momentum 
  line for an $S$=1/2 isotropic antiferromagnetic Heisenberg model on a square lattice of 100 sites ($L_x=L_y=10$) with 
  $\nod=2$ near the saturation field. 
  Note that the trivial component in the structure factors is subtracted for visibility (see the text). 
  The high symmetric momenta are indicated by $\Gamma$:~$(0,0)$, X:~$(\pi,0)$, and M:~$(\pi,\pi)$. 
  }
  \label{fig:spin_structures}
 \end{center}
\end{figure}

In Fig.~\ref{fig:spin_structures}, we also demonstrate the calculation of the static and dynamical spin structure factors for an $S=1/2$ isotropic antiferromagnetic Heisenberg model on a square lattice of 100 sites ($L_x=L_y=10$) with $\nod=2$. 
The static spin structure factor $S_{\bf q}^\alpha$ is defined as 
\begin{equation}
S_{\bf q}^\alpha = \bra{\phi} \hat{S}^{\alpha\dagger}_{{\bf q}} \hat{S}^{\alpha}_{{\bf q}} \ket{\phi}
\end{equation}
with $\ket{\phi}$ being the ground state and it is related to the dynamical spin structure factor $S_{\bf q}^\alpha(\omega)$ via
\begin{equation}
S_{\bf q}^\alpha = \int S_{\bf q}^\alpha(\omega)\, d\omega. 
\end{equation}
A nearly fully polarized state always displays a trivial but dominant sharp peak in the longitudinal structure factor at the $\Gamma$ point and the symmetrically equivalent momenta. 
For ease of visibility, this trivial component is subtracted form the calculated static and dynamical longitudinal spin 
structure factors, denoted respectively as $\tilde{S}_{\bf q}^z$ and $\tilde{S}_{\bf q}^z(\omega)$ in Fig.~\ref{fig:spin_structures}. 
Here, the dynamical spin structure factor $\tilde{S}_{\bf q}^z(\omega)$ at ${\bf q}={\bf 0}$ is given as 
\begin{equation}
\tilde{S}_{{\bf q}={\bf 0}}^z(\omega) \equiv {S}_{{\bf q}={\bf 0}}^z(\omega)-\frac{  \eta}{\pi (\omega^2+\eta^2)} \frac{M^2}{N} 
\end{equation}
and $\tilde{S}_{\bf q}^z(\omega)$ at ${\bf q}\ne{\bf 0}$ is exactly the same as ${S}_{\bf q}^z(\omega)$. 

\section{Summary}
\label{sec:summary}
We have developed the exact diagonalization package QS$^{3}$ for analyzing spin-1/2 quantum lattice models with XXZ interactions near the saturation field. 
The QS$^{3}$ package employs the symmetry-adapted basis sets with respect to the translational symmetry as well as the U(1) symmetry. 
In order to access large system sizes up to $O(1000)$, the QS$^{3}$ package does not use the traditional bit representation for spin configurations. 
Introducing OpenMP parallelization, the bottleneck of the calculation, i.e., large dimension matrix-vector multiplication, is efficiently accelerated by the parallelization. 
The QS$^{3}$ package computes fundamental physical quantities such as the local magnetization, two-point spin correlation function, and the dynamical spin structure factor. These quantities are essential and observable in experiments.
As demonstrated in the benchmark, the QS$^3$ package can treat three-dimensional systems to understand the ground state 
as well as the low-energy excitations with potentially interesting properties. 

For the future development, the QS$^{3}$ package will be extended to treat the point-group symmetry in addition to the translational symmetry. 
We will also introduce the multiple degrees of freedom per unit cell, i.e., multiple spins per unit cell, to treat more general lattice geometries such as the kagome and pyrochlore lattices.  
In addition, we will extend the application of the QS$^3$ packages to dilute fermionis, soft-core bosons, and higher-spin systems. 
These extensions are straightforward in terms of the coding employed in the QS$^{3}$ package and 
most likely increase a value of the QS$^3$ package as a research tool not only in condensed matter physics but also in quantum chemistry. For example, the QS$^3$ package will be able to handle the full configuration interaction (full CI) calculation for molecules with a small number of electrons occupying many orbitals, which are difficult to treat by an available open-source package, e.g., given in Ref.~\cite{Vogiatzis2017}. 

Furthermore, we can implement a function to simulate quantum circuits with symmetry constraints. 
This direction of development is important to provide reference data for benchmark results of future large-scale universal quantum computers and to investigate quantum accelerated algorithms for quantum many-body systems.  
These extensions are in part under progress and will be reported in the near future.

\section*{Acknowledgments}
H.U. thanks W. Mizukami for helpful comments. 
T.S. is supported by the Theory of Quantum Matter Unit of the Okinawa Institute
of Science and Technology Graduate University (OIST).
The work was partially supported by KAKENHI Nos.~17K14359, 18H01183, 19K14665, 21K03477, and 21H04446, 
and by JST PRESTO No.~JPMJPR1911.
This work was also supported by MEXT Q-LEAP Grant Number JPMXS0120319794. 
We are grateful for allocating computational resources of the HOKUSAI BigWaterfall supercomputing system at RIKEN. 
The QS$^3$ package was also developed and performed using the facilities of the Supercomputer Center, ISSP, 
the University of Tokyo, the facilities of the research center for computational science of national institutes of natural sciences, 
and the facilities of computing section, OIST.






\bibliographystyle{apsrev4-1}

\begin{thebibliography}{46}
\bibitem [{not()}]{note1}%
  \BibitemOpen
  \href@noop {} {}\bibinfo {note} {When we use standard 32 (64)-bit signed
  integer, the range is restricted from -2$^{31 (63)}$ to 2$^{31
  (63)}$+1.}\BibitemShut {Stop}%
\end{thebibliography}
%






\end{document}